\documentclass[
reprint,
amsmath,amssymb,
aps,
prb,
]{revtex4-1}

\usepackage{graphicx}
\usepackage{dcolumn}
\usepackage{bm}

\begin{document}

\preprint{APS/123-QED}

\title{Current noise of the interacting resonant level model} 

\author{T. J. Suzuki$^{1}$}

\author{D. M. Kennes$^{2}$}
\author{V. Meden$^2$}%
\affiliation{%
$^1$Department of Physics, University of Tokyo, Hongo, Tokyo 113-0033, Japan \\
$^2$Institut f\"{u}r Theorie der Statistischen Physik, RWTH Aachen University and JARA---Fundamentals of 
Future Information Technology, 52056 Aachen, Germany
}%

\date{\today}

\begin{abstract}
We study the zero-frequency current noise of the interacting resonant level model for arbitrary 
bias voltages using a functional renormalization group approach. For this we extend the existing 
nonequilibrium scheme by deriving and solving flow equations for the current-vertex functions. 
On-resonance artificial divergences of the latter found in lowest-order 
perturbation theory in the two-particle interaction are consistently removed. Away from 
resonance they are shifted to higher orders. This allows us to gain a comprehensive picture 
of the current noise in the scaling limit. At high bias voltages, the current noise exhibits 
a universal power-law decay, whose exponent is, to leading order in the interaction, identical 
to that of the current. The effective charge on resonance is analyzed in detail, 
employing properties of the vertex correction. We find that it is only modified to second or 
higher order in the two-particle interaction. 
\end{abstract}

\pacs{Valid PACS appear here}
\maketitle


\section{Introduction}


Remarkable advances in nanotechnology open up the possibility to explore transport 
beyond the linear response regime in experimentally well-controlled situations. 
Among nanoscale conductors, quantum dot systems have attracted much attention, as they offer a versatile arena in which to study nonequilibrium transport phenomena of interacting 
fermions. At sufficiently low temperatures, transport is dominated by quantum mechanics. 
In addition, the local two-particle interaction results in  fascinating many-body effects, 
often accompanied by the emergence of new energy scales. 
A generic nonequilibrium setup is given by a quantum dot coupled to several leads with different
chemical potentials.

The interacting resonant level model (IRLM) is a prominent example in which strong correlations play 
an essential role.\cite{PhysRevB.25.4815} It describes a single-level quantum dot dominated 
by charge fluctuations which are affected by the local two-particle interaction of dot and 
lead fermions. The IRLM was originally introduced as a close relative of the Kondo 
model,\cite{PhysRevB.25.4815} and, since then, many studies have been performed to elucidate 
its nonlinear 
transport properties.\cite{PhysRevLett.99.076806,PhysRevB.75.125107,boulat2008twofold,PhysRevLett.102.146803,karrasch2010functional,0295-5075-90-3-30003,PhysRevB.83.205103,PhysRevB.91.045140} These have 
shown that universal features appear in nonequilibrium transport if the lead bandwidth 
$\Delta$ is much larger than any other energy scale.
In this scaling regime, the current at large bias voltage is suppressed following a power law, 
whose exponent depends on the strength of the local interaction.\cite{PhysRevLett.99.076806,PhysRevB.75.125107,boulat2008twofold,karrasch2010functional,0295-5075-90-3-30003,PhysRevB.83.205103,PhysRevB.91.045140}


Accumulated knowledge in mesoscopic physics elucidates that the higher order cumulants of 
the current are of great importance to characterize the nonequilibrium transport.\cite{Blanter2001}
A vast amount of research on the current noise has shown that it contains information which 
cannot be obtained from  conductance measurements, e.g., the effective charge.\cite{saminadayar1997observation,*de1997direct,lefloch2003doubled,yamauchi2011evolution,ferrier2015universality}
While a unified picture for the current in the IRLM has been established by various methods, 
the understanding of its higher cumulants is rather limited for the moment.
A major obstacle is the absence of a general framework to treat the effects of strong 
correlations in a nonequilibrium situation.
To compute the current noise, one in general needs 
to consider the current-vertex function.\cite{PhysRevB.46.7061}
A perturbative approach to the current noise of the IRLM based on the
Keldysh technique was put forward in Ref.~\onlinecite{PhysRevB.76.193307}.
Important insights into the noise of the IRLM  under on-resonance conditions were gained by 
utilizing a special symmetry of this model for a particular interaction strength, 
which is known as self-duality.\cite{PhysRevB.52.8934} This symmetry makes it possible to 
map the IRLM to a solvable boundary sine-Gordon model even in the presence of a 
driving bias voltage.\cite{boulat2008twofold}
The effective charge of the quasiparticles of the IRLM at the self-dual point has been 
investigated using field-theoretical techniques and the density-matrix renormalization 
group method.\cite{branschadel2010shot,carr2011full,Carr2014} For the relatively 
strong interaction at which self-duality is established, the quasiparticles 
of the IRLM were found to have effective charge $e^{*}=2e$ by examining the shot 
noise,\cite{branschadel2010shot} 
which was confirmed computing the full counting statistics.\cite{carr2011full,2014arXiv1405.3070C}
This has to be contrasted to  $e^{*}=e$ in the noninteracting limit. In spite of this remarkable 
achievement, the bias voltage dependence of the current noise of the IRLM away from this self-dual 
point and off resonance is still an open question. In addition, finite temperature effects were so 
far not investigated. Considering this situation, it is strongly desirable to develop a systematic 
framework to calculate the current noise and its full counting statistics in general parameter 
regimes.


Recently, a functional renormalization group (FRG) method was developed to describe the 
nonequilibrium properties of correlated fermions.\cite{RevModPhys.84.299} Logarithmic divergences, 
which manifest themselves in plain perturbation theory even in the equilibrium IRLM, are consistently 
resummed employing FRG.\cite{karrasch2010functional} This method also contributed to deepen our 
understanding of the nonequilibrium transport properties of the 
IRLM.\cite{0295-5075-90-3-30003,karrasch2010functional} The renormalization of the hopping between 
the dot level and the leads can be described using a surprisingly simple 
approximation.\cite{karrasch2010functional} In this paper, we utilize the FRG approach to 
elucidate the current noise of the nonequilibrium IRLM. In FRG, we can obtain the 
vertex function by deriving and solving its flow equation.
For the level that is on resonance, an artificial divergence of the vertex correction obtained in lowest-order 
plain perturbation theory in the interaction is removed in the FRG scheme.
For off-resonance conditions, a severe divergence, found in perturbation theory 
if the level energy is aligned with one of the leads 
chemical potentials, is shifted to higher orders. These achievements allow us to gain a comprehensive picture 
of the zero-frequency current noise in the scaling limit.
We show that the current noise is governed by universal power-law scaling in the large bias 
voltage regime with an exponent which, to leading order in the interaction, is the same as that 
of the current. The effective charge is discussed by relating it to the vertex correction. We 
find that $e^{*} = e \left[ 1+ {\mathcal O}(u^2) \right] $ with the dimensionless amplitude 
of the interaction $u$.  


This paper is organized as follows. We describe the model and outline the FRG scheme to 
compute the current noise in Sec.~\ref{sec:model and formalism}. The results for the  
noise  are presented in Sec.~\ref{sec:result}. A summary is given in Sec.~\ref{sec:summary}.


\section{Model and formalism}
\label{sec:model and formalism}

\subsection{Model}
\label{subsec:model}

\begin{figure}[t]
\centering
\includegraphics[width=0.95\hsize]{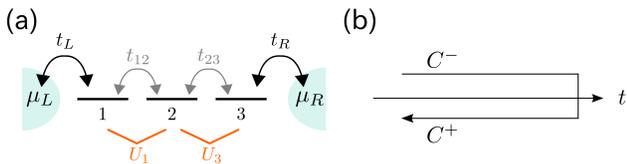}
\caption{(Color online)
(a) Schematic picture of the IRLM.
(b) The Keldysh contour.
}
\label{fig:System_Keldysh}
\end{figure}

The (two-reservoir) IRLM describes a spinless fermionic level coupled to 
delocalized fermions in the left and right 
leads. We consider the local repulsion between the fermion in this level and those in both leads.
This situation can be modeled by a three-site dot system, in which a particle occupying the central 
site feels the interaction with the adjacent ones. A schematic picture of the system is shown in 
Fig.~\ref{fig:System_Keldysh}(a). The hopping amplitudes $t_{L/R}>0$ between the three-site dot region 
and the leads are 
assumed to be much larger than the intersite ones, such that the lattice sites $1$ and $3$ are 
effectively incorporated into the leads.
The single-site model studied in the original paper\cite{PhysRevB.25.4815} is restored in this limit.\cite{karrasch2010functional,0295-5075-90-3-30003,jakobs2010properties,PhysRevB.81.195109}
The model is described by the action
\begin{align}
S
=&\sum^{3}_{i,j=1} \int d\tau d\tau' \bar{d}_{i}(\tau) {\bm g}^{-1}_{dij}(\tau,\tau')  d_{j}(\tau') \nonumber \\
&+\sum_{\alpha=L,R}\sum_{\bm{k}} \int d\tau d\tau' \bar{c}_{\alpha \bm{k}}(\tau) g^{-1}_{\alpha \bm{k}}(\tau,\tau') c_{\alpha \bm{k}}(\tau') \nonumber \\
&-\frac{1}{\sqrt{N}} \sum_{\bm{k}} \int d\tau
\Bigl[ t_L e^{iA_L(\tau)}\bar{d}_{1}(\tau)c_{L\bm{k}}(\tau)  \nonumber \\
&\hspace{40pt} +  t_R \bar{d}_{3}(\tau)c_{R\bm{k}}(\tau) + {\rm H.c.} \Bigr]  + S_{U},
\label{eq1}
\end{align}
where the isolated lead and dot Green's functions are given by 
$ g^{-1}_{\alpha\bm{k}}(\tau,\tau') \equiv \left[ i\frac{d}{d\tau'} - \epsilon_{\alpha \bm{k}} \right] \delta(\tau,\tau')$ and 
\begin{widetext}
\begin{align}
{\bm g}^{-1}_{d}(\tau,\tau')=
\left(
    \begin{array}{ccc}
      i\frac{d}{d\tau'} -(\epsilon_{1} - U_1/2) & -t_{12}            & 0  \\
      -t_{12}       & i\frac{d}{d\tau'} - (\epsilon_{2} -(U_1+U_3)/2)  & -t_{23}\\
      0          & -t_{23}         & i\frac{d}{d\tau'} - (\epsilon_{3} - U_3/2)
    \end{array}
  \right) \delta(\tau,\tau'),
\end{align}
\end{widetext}
respectively, and the interaction part is given as
\begin{align}
  S_{U}
& \equiv - \int d\tau \left( U_1 \bar{d}_{2}(\tau)d_{2}(\tau)\bar{d}_{1}(\tau)d_{1}(\tau) \right. \nonumber \\
&\hspace{40pt} \left.+  U_3 \bar{d}_{2}(\tau)d_{2}(\tau)\bar{d}_{3}(\tau)d_{3}(\tau) \right).
\end{align}
The argument $\tau$ combines the time $t$ and the Keldysh index $\nu=\mp$
[see Fig.~\ref{fig:System_Keldysh}(b)]. 
The symbol $\int d\tau $ denotes integration over $t$ and summation over $\nu$.    
The Grassmann field $\bar{d}_{i}(\tau)$ $[d_{i}(\tau)]$ creates [annihilates] a spinless 
fermion at time $t$, on the Keldysh contour branch with index $\nu$, and on lattice 
site {\it i}. Similarly  
$\bar{c}_{\alpha {\bm k}}(\tau)$ $[c_{\alpha {\bm k}}(\tau)]$ is a Grassmann field for creating [annihilating]
a fermion in lead $\alpha$ with momentum ${\bm k}$. The Green's functions in  Eq.~(\ref{eq1}) 
must be understood as $2 \times 2$ matrices in the Keldysh space and the addends contain 
matrix products which are left implicit.
The delta function on the Keldysh contour is defined as $\delta(\tau,\tau') \equiv \delta(t-t') \sigma^{\nu \nu'}_{z}$ with the standard Pauli matrix $\sigma_{z}$.  
The energy level of each site is denoted by $\epsilon_{i}$ for $i=1,2,3$, and the hopping 
amplitude between the sites 1 (3) and 2 by $t_{12}>0$ $(t_{23}>0)$. The energy level $\epsilon_{i}$ is 
defined such that $\epsilon_1=\epsilon_2=\epsilon_3=0$ corresponds to the particle-hole 
symmetric case. The third term of the action Eq.~(\ref{eq1})
describes the hopping between the quantum dot and 
the leads with $N$ sites. In our expressions we always take the thermodynamic 
limit $N \to \infty$. The auxiliary vector potential $A_L(\tau)$ is incorporated using 
the Peierls substitution and later used as a source field to generate the current-vertex 
function.\cite{kamenev2011field} We choose units with $k_B=1$, $\hbar=1$, and elementary charge 
$e=1$.

\subsection{Generating functional}
\label{subsec:generating functional}

The noninteracting reservoirs can be integrated out yielding the action
\begin{align}
S
=&\sum^{3}_{i,j=1} \int d\tau d\tau' \bar{d}_{i}(\tau) {\bm G}^{-1}_{0ij}(\tau,\tau')  d_{j}(\tau') 
+ S_{U},
\end{align}
where $ {\bm G}^{-1}_{0}(\tau,\tau') \equiv {\bm g}^{-1}_{d}(\tau,\tau') - {\bm \Sigma}^{-1}_{\rm res}(\tau,\tau') $.
The tunneling self-energy is given as
\begin{align}
\label{eq:tunneling self-energy for IRLM}
&\bm{\Sigma}_{\rm res}(\tau,\tau') \nonumber \\
&= 
\left(
    \begin{array}{ccc}
      t^2_Le^{i[A_L(\tau)-A_L(\tau')]} g_{L}(\tau,\tau') & 0 & 0  \\
      0  & 0 & 0 \\
      0  & 0 & t^2_R g_{R}(\tau,\tau')
    \end{array}
  \right),
\end{align}
with $ g_{\alpha}(\tau,\tau') = \frac{1}{N}\sum_{{\bm k}}g_{\alpha{\bm k}}(\tau,\tau')$.
We consider the system with a time-independent bias voltage $V$, which is applied symmetrically 
to the leads, i.e., $\mu_L=V/2$ and $\mu_R=-V/2$. Both of the leads are assumed to be in equilibrium.
Since in its bias-voltage-driven nonequilibrium steady state the system is invariant under 
timetranslation, we perform a Fourier transform to energy representation. 

As usual in the Keldysh formalism we use the representation with the retarded (r), Keldysh (K), and advanced (a) components instead of the one with $\nu=\mp$ if appropriate.
The component which vanishes in the two-point Green's function due to causality but appears for other vertices (see below) is denoted by $\tilde{\mbox K}$.    
If the source field $A_L(\tau)$ is set to be zero, the relevant 
parts of the tunneling self-energy are obtained as
\begin{align}
\bm{\Sigma}^{\rm r}_{\rm res}(\omega) &=
\left(
    \begin{array}{ccc}
      \displaystyle   - \frac{i\Delta_L}{2} & 0 & 0  \\
      0  & 0 & 0 \\
      0  & 0 &  \displaystyle - \frac{i\Delta_R}{2}
    \end{array}
  \right), \\
\bm{\Sigma}^{\rm K}_{\rm res}(\omega) &=
\left(
    \begin{array}{ccc}
      i \Delta_L \left(2f_L(\omega)-1\right) & 0 & 0  \\
      0  & 0 & 0 \\
      0  & 0 & i \Delta_R \left( 2f_R(\omega) -1 \right)
    \end{array}
  \right).
\end{align}
Here, we use the wide-band limit and define the bandwidth, $\Delta_{\alpha} \equiv 2\pi 
\rho_{\alpha} t_{\alpha}^2$. For simplicity, we assume $\Delta_L=\Delta_R \equiv \Delta$, 
$t_{12}=t_{23}\equiv t$, $\epsilon_1 = \epsilon_3=0$, $\epsilon_2=\epsilon$, and 
$U_1=U_3\equiv U \geq 0$. This implies that at particle-hole symmetry $\epsilon=0$ 
transport is resonant. We thus refer to this case either as the particle-hole symmetric 
point or the on-resonance situation.

\begin{figure}[b!]
\centering
\includegraphics*[width=0.8\hsize]{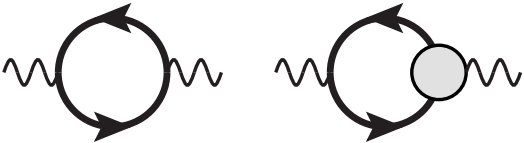}
\caption{
  \label{fig:Diagram of Current noise}
  Diagrammatic representation of the current noise.
  The circle represents the three-point vertex function.
  The left diagram is called the bubble term, while the right is called the vertex correction.
}
\end{figure}

Following the standard functional integral approach to quantum many-body 
physics,\cite{kamenev2011field} we introduce the additional source term
\begin{align}
S^{\rm s} &\equiv \sum^{3}_{i=1} \int d\tau \left[ 
\bar{\eta}_i(\tau)d_i(\tau) +\bar{d}_i(\tau)\eta_i(\tau) \right],
\end{align}
which allows us to generate correlation functions by functional derivatives. 
The corresponding generating functional is given by
\begin{align}
W[\eta,\bar{\eta},A] \equiv -i \ln \int {\cal D} \left[ \bar{d},d \right] \exp \left[i (S + S^{\rm s}) \right].
\end{align}
The effective action\cite{CHOU19851} is defined using the Legendre transform as
\begin{align}
\Gamma [ \langle d \rangle^{\rm s}, \langle \bar{d} \rangle^{\rm s}, A]
&\equiv W [\eta,\bar{\eta},A]
- \sum^{3}_{i=1}\int d\tau \left[ \bar{\eta}_{i}(\tau) \langle d_{i} \rangle^{\rm s} (\tau)  \right. \nonumber \\
&\hspace{80pt} \left.+ \langle \bar{d}_{i} \rangle^{\rm s} (\tau) \eta_{i}(\tau) \right],
\end{align}
with
\begin{align}
\langle {\cal O} \rangle ^{\rm s} \equiv  \frac{\int {\cal D} \left[ \bar{d}, d \right] {\cal O} \exp \left[i (S+S^{\rm s} ) \right] }{\int {\cal D} \left[ \bar{d}, d \right] \exp \left[i (S+S^{\rm s} ) \right]} .
\end{align}
This effective action acts as the generating functional of one-particle irreducible vertex functions
(e.g.,~the self-energy). The vertex expansion of the effective action is given as 
\begin{widetext}
\begin{align}
  \Gamma [ \langle d \rangle^{\rm s}, \langle \bar{d} \rangle^{\rm s}, A] 
  =  \sum^{\infty}_{m,n=0}\frac{(-1)^m}{(m!)^2}\frac{1}{n!} 
  & \prod^{m}_{j,k=0}\prod^{n}_{l=0} \int d\tau_j d\tau'_k d\tau''_l 
 \; \gamma^{(2m,n)}_{i'_1\cdots i'_m;i_1\cdots i_m;\alpha_1\cdots \alpha_n}(\tau'_1,\cdots,\tau'_m; \tau_1,\cdots,\tau_m; 
  \tau''_1,\cdots,\tau''_n) 
 \nonumber \\ & \times
  \langle \bar{d}_{i'_1} \rangle^{\rm s}(\tau'_1) \cdots \langle \bar{d}_{i'_m} \rangle^{\rm s}(\tau'_m) 
\langle d_{i_1} \rangle^{\rm s}(\tau_1) \cdots \langle d_{i_m} \rangle^{\rm s}(\tau_m) A_{\alpha_1}(\tau''_1) \cdots A_{\alpha_n}(\tau''_n),
\end{align}
where repeated site indices are summed over.
We note that the auxiliary vector potential $A_{\alpha}(\tau)$ is also defined on the Keldysh contour;
i.e., it has the two components $A^{-}_{\alpha}(t)$ and $A^{+}_{\alpha}(t)$.
Hence, there are in total $2^{2m+n}$ components for the current-vertex function 
$\gamma^{(2m,n)}$.

The doubled degrees of freedom of the vector potential allow one to completely describe both the 
dynamical evolution as well as the statistical correlation.\cite{kamenev2011field,CHOU19851}
We define the source and the physical component of the vector potential as
\begin{align}
A^{\rm s}_{\alpha}(t)\equiv \frac{1}{2}\left[ A^{-}_{\alpha}(t) - A^{+}_{\alpha}(t) \right], \quad
A^{\rm p}_{\alpha}(t)\equiv \frac{1}{2}\left[ A^{-}_{\alpha}(t) + A^{+}_{\alpha}(t) \right],
\end{align}
respectively.
We utilize $A^{\rm s}(t)$ to derive the expression of the current-vertex functions and 
their flow equations.
After deriving all the equations, we take the physically relevant limit $A^{\rm s}(t)\to 0$.
As we introduced the bias voltage via the chemical potentials of the leads and 
do not consider any further fields, the physical component of the vector potential 
is set to zero as well. 
The source and physical components of the current-vertex functions are defined as
\begin{align}
\left(\gamma^{(0,1)}\right)^{\rm s}_{\alpha} (t) 
\equiv \left. \frac{\delta \Gamma}{\delta A^{\rm s}_{\alpha}(t)} \right|_{A_\alpha^{\mp}=0},  \quad
\left(\gamma^{(0,1)}\right)^{\rm p}_{\alpha} (t)
\equiv \left. \frac{\delta \Gamma}{\delta A^{\rm p}_{\alpha}(t)} \right|_{A_\alpha^{\mp}=0} ,
\end{align}
respectively.
The components of the higher order current-vertex functions are defined in the same way.
Then, the current noise can be written as\cite{PhysRevB.87.235303}
\begin{align}
  S_{\alpha\alpha'}(t,t') &\equiv 
  \langle I_{\alpha}(t)I_{\alpha'}(t')  \rangle
  +\langle I_{\alpha'}(t')I_{\alpha}(t)  \rangle
  -2\langle I_{\alpha}(t) \rangle \langle I_{\alpha'}(t')  \rangle \nonumber \\
  &=S^{0}_{\alpha\alpha'}(t,t') + S^{U}_{\alpha\alpha'}(t,t'),  \label{eq:current noise}
\end{align}
where the two terms are defined as
\begin{align}
\label{eq:expression of bubble term}
S^{0}_{\alpha\alpha'}(t,t') 
&\equiv \frac{1}{2}  \int dt_1 dt'_1 \bm{G}^{\nu_1\nu'_1}_{i_1 i'_1}(t_1,t'_1)  \sigma^{\nu'_1\nu'_1}_{z} \left( \bm{\gamma}^{(2,2)}_{\rm res}  \right)^{\nu'_1\nu_1;{\rm ss}}_{i'_1i_1;\alpha\alpha'}(t'_1,t_1;t,t')  \sigma^{\nu_1\nu_1}_{z}\nonumber \\
& \hspace{20pt} + \frac{1}{2}  \int dt_1 dt_2 dt'_1 dt'_2 \bm{G}^{\nu_2\nu'_1}_{i_2 i'_1}(t_2,t'_1)  \sigma^{\nu'_1\nu'_1}_{z} \left( \bm{\gamma}^{(2,1)}_{\rm res} \right)^{\nu'_1\nu_1;{\rm s}}_{i'_1i_1;\alpha}(t'_1,t_1;t) \sigma^{\nu_1\nu_1}_{z} \nonumber \\
& \hspace{120pt} \times \bm{G}^{\nu_1\nu'_2}_{i_1 i'_2} (t_1,t'_2) \sigma^{\nu'_2\nu'_2}_{z} \left( \bm{\gamma}^{(2,1)}_{\rm res} \right)^{\nu'_2\nu_2;{\rm s}}_{i'_2i_2;\alpha'}(t'_2,t_2;t')\sigma^{\nu_2\nu_2}_{z}, \\
\label{eq:expression of vertex correction}
S^{U}_{\alpha\alpha'}(t,t') 
&\equiv \frac{1}{2}  \int dt_1 dt_2 dt'_1 dt'_2 
  \bm{G}^{\nu_2\nu'_1}_{i_2 i'_1}(t_2,t'_1)  \sigma^{\nu'_1\nu'_1}_{z} \left( \bar{\bm{\gamma}}^{(2,1)} \right)^{\nu'_1\nu_1;{\rm s}}_{i'_1i_1;\alpha}(t'_1,t_1;t) \sigma^{\nu_1\nu_1}_{z} \nonumber \\
& \hspace{100pt}  \times \bm{G}^{\nu_1\nu'_2}_{i_1 i'_2} (t_1,t'_2) \sigma^{\nu'_2\nu'_2}_{z} \left( \bm{\gamma}^{(2,1)}_{\rm res} \right)^{\nu'_2\nu_2;{\rm s}}_{i'_2i_2;\alpha'}(t'_2,t_2;t')\sigma^{\nu_2\nu_2}_{z},
\end{align}
with
\begin{align}
  \left( \bar{\bm{\gamma}}^{(2,1)} \right)^{\nu'_1\nu_1;{\rm s}}_{i'_1i_1;\alpha}(t'_1,t_1;t) 
\equiv
  \left( \bm{\gamma}^{(2,1)} \right)^{\nu'_1\nu_1;{\rm s}}_{i'_1i_1;\alpha}(t'_1,t_1;t)
  -\left( \bm{\gamma}^{(2,1)}_{\rm res} \right)^{\nu'_1\nu_1;{\rm s}}_{i'_1i_1;\alpha}(t'_1,t_1;t).
\label{eq:gammabar}
\end{align}
\end{widetext}
Here, $\gamma^{(2,n)}_{\rm res}$ (for $n>0$) is the noninteracting part of the $(2+n)$-point 
current-vertex function. 
Thus $ \bar{\bm{\gamma}}^{(2,1)}$ defined in Eq.~(\ref{eq:gammabar}) is the interaction induced 
part of the three-point vertex function.   
In Eqs.~(\ref{eq:expression of bubble term}) and (\ref{eq:expression of vertex correction}) 
the repeated Keldysh indices $\nu=\mp$ and site indices are summed over.
The first term $S^0$ on the right-hand side 
of Eq.~(\ref{eq:current noise}) is called the bubble term, while the second $S^U$ is the vertex 
correction to the noise. 
Their diagrammatic representations are shown in Fig.~\ref{fig:Diagram of Current noise}.
The effect of the repulsive interaction is included in the self-energy of the propagator, ${\bm \Sigma}_{U}= {\bm G}^{-1}_{0}  - {\bm \gamma}^{(2,0)}$, and the three-point vertex function, $\gamma^{(2,1)}$, both being determined by the FRG approach.

\subsection{Functional renormalization group approach}
\label{subsec:FRG}

In setting up the functional renormalization group approach,\cite{RevModPhys.84.299} we use a
reservoir cutoff as the flow parameter.\cite{jakobs2010properties,PhysRevB.81.195109} The Kubo-Martin-Schwinger (KMS) 
condition is automatically preserved in this scheme, which is important to perform  
calculations consistent with the fluctuation dissipation theorem in the limit $V \rightarrow0$.
The flow parameter is introduced as an additional tunneling self-energy,
\begin{align}
\bm{\Sigma}^{\rm r}_{{\rm aux},\Lambda}(\omega) &= - \frac{i\Lambda }{2} \bm{1}, \\
\bm{\Sigma}^{\rm K}_{{\rm aux},\Lambda}(\omega) &=i \Lambda \left[2f_{\rm aux}(\omega) -1 \right] \bm{1},
\end{align}
where $f_{\rm aux}(\omega)$ is the Fermi-Dirac distribution function of the auxiliary structureless 
reservoirs and $\bm{1}$ is the identity matrix of dimension $3$.
It was examined earlier that results for the current are independent of the choice of the 
temperature in the auxiliary reservoirs.\cite{PhysRevB.85.085113,PhysRevB.87.075130}
In this paper, we utilize the auxiliary reservoirs with infinite temperature and thus 
$f_{\rm aux}(\omega)=1/2$.
This simplifies the flow equations as the Keldysh component of the auxiliary self-energy 
vanishes. The full Green's function is obtained by the Dyson equation
\begin{align}
\left( \bm{G}^{\rm r}_{\Lambda} \right) ^{-1}(\omega)
&= \left( \bm{G}^{\rm r}_{0} \right) ^{-1}(\omega) - \bm{\Sigma}^{\rm r}_{{\rm aux},\Lambda}(\omega) - 
\bm{\Sigma}^{\rm r}_{U,\Lambda}(\omega), \\
\left( \bm{G}^{\rm K}_{\Lambda} \right) (\omega)
&= \bm{G}^{\rm r}_{\Lambda} (\omega) \left[\bm{\Sigma}^{\rm K}_{\rm res} (\omega) + \bm{\Sigma}^{\rm K}_{U,\Lambda} (\omega) \right] \bm{G}^{\rm a}_{\Lambda} (\omega) .
\end{align}
The scale-dependent propagator $\bm{S}_{\Lambda}(\tau,\tau')$ appearing in the RG flow equations 
(see below) is defined as
\begin{align}
\bm{S}_{\Lambda}(\tau,\tau') \equiv \int d\tau_1 d\tau_2 \bm{G}_{\Lambda}(\tau,\tau_1)
\frac{d \bm{\Sigma}_{{\rm aux},\Lambda}(\tau_1,\tau_2)}{d\Lambda}\bm{G}_{\Lambda}(\tau_2,\tau').
\end{align}
with components
\begin{align}
\bm{S}^{\rm r}_{\Lambda}(\omega) 
&= \frac{-i}{2} \bm{G}^{\rm r}_{\Lambda}(\omega) \bm{G}^{\rm r}_{\Lambda}(\omega), \\
\bm{S}^{\rm K}_{\Lambda}(\omega) 
&= \frac{-i}{2}  \bm{G}^{\rm r}_{\Lambda}(\omega) \bm{G}^{\rm K}_{\Lambda}(\omega) 
+ \frac{i}{2} \bm{G}^{\rm K}_{\Lambda}(\omega) \bm{G}^{\rm a}_{\Lambda}(\omega).
\end{align}

We consider the model with $\Lambda_{\rm init} \rightarrow \infty$ as the initial 
one of the flow, as all the 
vertex functions can be  calculated exactly in this limit. The initial conditions of the self-energy 
and the vertex functions are summarized in Appendix~\ref{app:Initial conditions}. The set of 
coupled flow equations has to be integrated down to $\Lambda=0$, at which the auxiliary reservoirs are 
decoupled and the cutoff-free problem of interest is restored. 

\begin{figure}[t]
\centering
\includegraphics*[width=\hsize]{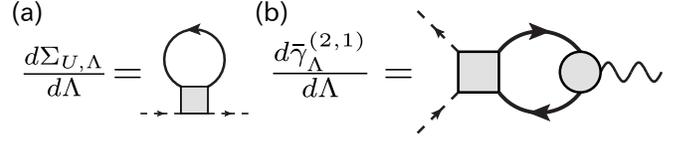}
\caption{
\label{fig:Flow_eq}
Diagrammatic representation of the flow equations of (a) the self-energy and (b) the three-point current-vertex function in the static approximation.
The square represents the two-particle interaction.
}
\end{figure}

In order to implement numerical calculations, we need to truncate the infinite hierarchy of the flow 
equations to a given order. In this paper, we use the lowest order truncation, which is known as 
the static approximation,\cite{RevModPhys.84.299} to determine the flow equations of the self-energy 
and current-vertex functions. Flow equations are truncated at the first order in the interaction, 
$U$. The remaining terms are the Hartree-Fock-type diagram for the self-energy -- we note that 
due to the underlying RG procedure our approximation is not equivalent to the Hartree-Fock approximation -- and 
the RPA-type diagram for the  vertex function. The diagrammatic representation of the flow equations 
is given in Fig.~\ref{fig:Flow_eq}. In spite of this simple treatment, this approximation for the self-energy 
is known to describes the rich properties of nonequilibrium transport due to the built-in 
renormalization.\cite{RevModPhys.84.299} In particular, logarithmic divergences 
found in the scaling limit of the IRLM in plain perturbation theory are consistently resummed to 
power laws.\cite{karrasch2010functional}  The current-vertex function has not yet been treated 
within the present truncated FRG scheme, and its role for the current noise is discussed below.
We note that higher order corrections can be systematically included in principle by incorporating 
flow equations of higher order vertices.

The flow equation of the four-point vertex function is ignored in the static approximation, and 
its value is replaced by the initial one which is given by the 
anti-symmetrized bare two-particle interaction $U_{ik;jl}$ (see Appendix~\ref{app:Initial conditions}).

The flow equation of the self-energy is obtained as
(repeated site indices are summed over)
\begin{align}
\frac{d }{d\Lambda}\left({\bm \Sigma}^{\rm r}_{U,\Lambda}\right)_{ij}
\label{eq:flow eq. of the self-energy}
= \frac{iU_{ik;jl}}{2} \int \frac{d\omega}{2\pi} \left(\bm{S}^{\rm K}_{\Lambda}\right)_{lk} (\omega) .
\end{align}
Within the present approximation, the self-energy is frequency independent due to the structure of 
the right-hand side. Hence, single-particle Green's functions can be interpreted as effective 
noninteracting ones with renormalized parameters.

The flow equation of the retarded component of the interaction-induced 
part of the three-point vertex function [see Eq.~(\ref{eq:gammabar})] is given by
\begin{align}
\frac{d}{d\Lambda} \left( \bar{{\bm \gamma}}^{(2,1)}_{\Lambda} \right)^{{\rm r};{\rm s}}_{ij;L}
\label{eq:flow eq. of ret. three-point vertex func.}
= \frac{iU_{ik;jl}}{2} 
\left[
  \left(\bm{\Phi}^{(2,1)}_{\Lambda}\right)^{\rm K}_{lk}
  +\left(\bm{\Phi}^{(2,1)}_{\Lambda}\right)^{\rm \tilde{K}}_{lk} 
\right]
,
\end{align}
with
\begin{align}
&\left( \bm{\Phi}^{(2,1)}_{\Lambda} \right)^{\rm K} \equiv \int \frac{d\omega}{2\pi}
\left[ 
  {\bm S}^{\rm r}_{\Lambda}(\omega) \left( {\bm \gamma}^{(2,1)}_{\Lambda} \right)^{{\rm r};{\rm s}}_{;L} {\bm G}^{\rm K}_{\Lambda}(\omega)  \right. \nonumber \\
& +  {\bm S}^{\rm K}_{\Lambda}(\omega) \left( {\bm \gamma}^{(2,1)}_{\Lambda} \right)^{{\rm a};{\rm s}}_{;L} {\bm G}^{\rm a}_{\Lambda}(\omega)
+  {\bm S}^{\rm r}_{\Lambda}(\omega) \left( {\bm \gamma}^{(2,1)}_{\Lambda} \right)^{{\rm K};{\rm s}}_{;L} {\bm G}^{\rm a}_{\Lambda}(\omega) \nonumber \\
& \left.
+  {\bm S}^{\rm K}_{\Lambda}(\omega) \left( {\bm \gamma}^{(2,1)}_{\Lambda} \right)^{\tilde{{\rm K}};{\rm s}}_{;L} {\bm G}^{\rm K}_{\Lambda}(\omega)
+({\bm S} \leftrightarrow {\bm G})
\right],
\end{align}
and
\begin{align}
&\left( \bm{\Phi}^{(2,1)}_{\Lambda} \right)^{\rm {\tilde K}} \nonumber \\
&\equiv \int \frac{d\omega}{2\pi}
\left[
{\bm S}^{\rm a}_{\Lambda}(\omega) \left( {\bm \gamma}^{(2,1)}_{\Lambda} \right)^{\tilde{{\rm K}};s}_{;L} {\bm G}^{\rm r}_{\Lambda}(\omega)
+({\bm S} \leftrightarrow {\bm G})
  \right].
\end{align}
The abbreviation $({\bm S} \leftrightarrow {\bm G})$ denotes the terms which are obtained by 
mutually replacing ${\bm S}$ and ${\bm G}$ in the preceding ones in the same parenthesis.
Similarly, the flow equation of the Keldysh component is obtained as
\begin{align}
\frac{d}{d\Lambda} \left( \bar{{\bm \gamma}}^{(2,1)}_{\Lambda} \right)^{{\rm K};{\rm s}}_{ij;L}
\label{eq:flow eq. of Keldysh three-point vertex func.}
= \frac{iU_{ik;jl}}{2} 
\left[
  \left(\bm{\Phi}^{(2,1)}_{\Lambda}\right)^{\rm r}_{lk}
  +\left(\bm{\Phi}^{(2,1)}_{\Lambda}\right)^{\rm a}_{lk} 
\right]
.
\end{align}
with
\begin{align}
&\left( \bm{\Phi}^{(2,1)}_{\Lambda} \right)^{\rm r}
\equiv \int \frac{d\omega}{2\pi}
\left[ 
  {\bm S}^{\rm r}_{\Lambda}(\omega) \left( {\bm \gamma}^{(2,1)}_{\Lambda} \right)^{{\rm r};{\rm s}}_{;L} {\bm G}^{\rm r}_{\Lambda}(\omega) \right. \nonumber \\
  &\left.
  +  {\bm S}^{\rm K}_{\Lambda}(\omega) \left( {\bm \gamma}^{(2,1)}_{\Lambda} \right)^{\tilde{{\rm K}};{\rm s}}_{;L} {\bm G}^{\rm r}_{\Lambda}(\omega)
+ ({\bm S} \leftrightarrow {\bm G})
  \right],
\end{align}
and
\begin{align}
&\left( \bm{\Phi}^{(2,1)}_{\Lambda} \right)^{\rm a}
\equiv \int \frac{d\omega}{2\pi}
\left[ 
{\bm S}^{\rm a}_{\Lambda}(\omega) \left( {\bm \gamma}^{(2,1)}_{\Lambda} \right)^{{\rm a};{\rm s}}_{;L} {\bm G}^{\rm a}_{\Lambda}(\omega) \right. \nonumber \\
& \left.
 +  {\bm S}^{\rm a}_{\Lambda}(\omega) \left( {\bm \gamma}^{(2,1)}_{\Lambda} \right)^{\tilde{{\rm K}};{\rm s}}_{;L} {\bm G}^{\rm K}_{\Lambda}(\omega)
+ ({\bm S} \leftrightarrow {\bm G})
  \right].
\end{align}
Again, repeated site indices are summed over.
The argument of the three-point vertex functions is omitted as these turn out to be independent of 
frequency in the static approximation. In contrast to the self-energy, these vertex functions do not 
have a simple interpretation in reference to a noninteracting model.

Using the initial condition and the flow equation, we can prove that the three-point current-vertex 
functions fulfill the symmetry relations
\begin{align}
\left[ \left( {\bm \gamma}^{(2,1)}_{\Lambda} \right)^{{\rm r};{\rm s}}_{ij;L} \right]^{*}
\label{eq:symmetry relation 4 of current-vertex functions}
&= -\left( {\bm \gamma}^{(2,1)}_{\Lambda} \right)^{{\rm a};{\rm s}}_{ji;L}, \\
\left[ \left( {\bm \gamma}^{(2,1)}_{\Lambda} \right)^{{\rm K};{\rm s}}_{ij;L} \right]^{*}
\label{eq:symmetry relation 5 of current-vertex functions}
&= \left( {\bm \gamma}^{(2,1)}_{\Lambda} \right)^{{\rm K};{\rm s}}_{ji;L}, \\
\left[\left( {\bm \gamma}^{(2,1)}_{\Lambda} \right)^{{\rm \tilde{K}};{\rm s}}_{ij;L}\right]^{*}
\label{eq:symmetry relation 6 of current-vertex functions}
&= \left( {\bm \gamma}^{(2,1)}_{\Lambda} \right)^{{\rm {\tilde{K}}};{\rm s}}_{ji;L}.
\end{align}
In the static approximation, we can derive the additional relations
\begin{align}
  \left( \bar{{\bm \gamma}}^{(2,1)}_{\Lambda} \right)^{{\rm r};{\rm s}}_{ij;L}
  &= \left( \bar{{\bm \gamma}}^{(2,1)}_{\Lambda} \right)^{{\rm a};{\rm s}}_{ij;L}, \\
  \left( \bar{{\bm \gamma}}^{(2,1)}_{\Lambda} \right)^{{\rm K};{\rm s}}_{ij;L}
  &= \left( \bar{{\bm \gamma}}^{(2,1)}_{\Lambda} \right)^{{\rm {\tilde{K}}};{\rm s}}_{ij;L}.
\end{align}
Hence, it is sufficient to determine the retarded and the Keldysh component of the three-point vertex function.

We determine the self-energy and the three-point current-vertex functions by solving these 
flow equations numerically, and use the $\Lambda=0$ functions in the formula of the current 
noise given in Eqs.~(\ref{eq:current noise})-(\ref{eq:gammabar}). 
Details of the numerical implementation are 
given in Appendix~\ref{app:numerical details}.

\section{Results}
\label{sec:result}

\subsection{A consistency check for the current}
\label{subsec:consistency with MW}

There have already been an extensive number of studies on the steady-state current of the IRLM 
to elucidate its rich properties.\cite{PhysRevLett.99.076806,PhysRevB.75.125107,boulat2008twofold,PhysRevLett.102.146803,karrasch2010functional,0295-5075-90-3-30003,PhysRevB.83.205103,PhysRevB.91.045140}
We here focus on the case of the bandwidth $\Delta$ being much larger than the other energy scales of the 
problem. It is known as the scaling limit, and universal features of the steady-state 
current, such as the power-law behavior at large bias voltages, manifest 
themselves.\cite{PhysRevLett.99.076806,PhysRevB.75.125107,boulat2008twofold,karrasch2010functional,0295-5075-90-3-30003,PhysRevB.83.205103,PhysRevB.91.045140}
In an earlier FRG approach, the current was computed from the self-energy employing the Meir-Wingreen 
formula.\cite{PhysRevLett.68.2512} In this subsection, it is discussed that an alternative FRG formulation 
can be developed in which a flow equation for the current is derived and solved.
We show that both schemes provide the same results up to linear order in $U$, which is 
the one to which our truncation is controlled. As the flowing current-vertex functions derived above enter the flow equation for the current this provides 
a nontrivial consistency-check of our formulation later used to study the noise.

\begin{figure}[t]
\centering
\includegraphics*[width=0.7\hsize]{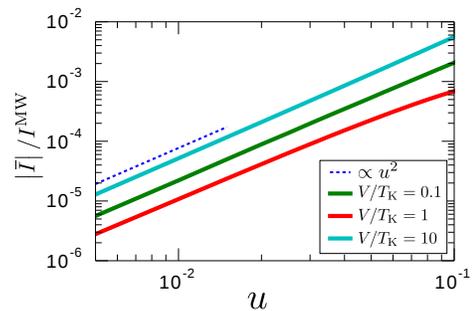}
\caption{
  \label{fig:Diff_Current_vs_U} 
  (Color online)
  The dependence of $\bar{I}$ on the dimensionless interaction $u$ for various $V$.
  The unit of energy, $T_{\rm K}$, is introduced in Sec.~\ref{subsec:noise in the scaling limit}.
  The parameters are $t / \Delta=0.001$, $\epsilon / T_{\rm K}=0$, and $T / T_{\rm K}=0$.
}
\end{figure}

The explicit expression of the current using the Meir-Wingreen formula is given as
\begin{align}
  I^{\rm MW}_{\Lambda} = \frac{1}{2\pi} \int d\omega 
&\left[
T^{\Lambda}_{LR}(\omega)(f_L(\omega)-f_R(\omega)) \right. \nonumber \\
& \left. +T^{\Lambda}_{L {\rm aux}}(\omega)(f_L(\omega)-f_{\rm aux}(\omega))
\right],
\end{align}
with
\begin{align}
T^{\Lambda}_{LR}(\omega)=
& \Delta_L \Delta_R \left( {\bm G}^{\rm r}_{\Lambda} \right)_{13}(\omega) \left( {\bm G}^{\rm a}_{\Lambda} \right)_{31}(\omega), \\
T^{\Lambda}_{L {\rm aux}}(\omega)=
& \Delta_L \Lambda \left( {\bm G}^{\rm r}_{\Lambda} {\bm G}^{\rm a}_{\Lambda}\right)_{11}(\omega).
\end{align}
This should be equivalent to the current obtained by solving its flow equation, which is denoted 
by $I^{\rm flow}_{\Lambda}$.
If we focus on their difference, i.e., $ \bar{I}_{\Lambda} \equiv I^{\rm MW}_{\Lambda}-I^{\rm flow}_{\Lambda}$, 
its flow equation is given by
\begin{align}
\frac{d \bar{I}_{\Lambda} }{d\Lambda}
= &\frac{-i}{2} \int \frac{d\omega}{2\pi} \left( {\bm S}_{\Lambda} \right)^{\rm r}_{ij}(\omega)  \left( \bar{{\bm \gamma}}^{(2,1)}_{\Lambda} \right)^{{\rm r};{\rm s}}_{ji;L}(\omega;\omega;0) \nonumber \\
&  - \frac{i}{2} \int \frac{d\omega}{2\pi} \left( {\bm S}_{\Lambda} \right)^{\rm a}_{ij}(\omega)  \left( \bar{{\bm \gamma}}^{(2,1)}_{\Lambda} \right)^{{\rm a};{\rm s}}_{ji;L}(\omega;\omega;0) \nonumber \\
&  - \frac{i}{2} \int \frac{d\omega}{2\pi} \left( {\bm S}_{\Lambda} \right)^{\rm K}_{ij}(\omega)  \left( \bar{{\bm \gamma}}^{(2,1)}_{\Lambda} \right)^{\tilde{\rm K};{\rm s}}_{ji;L}(\omega;\omega;0) \nonumber \\
& + \frac{i}{2} \Delta_L \int \frac{d\omega}{2\pi}
\left( (2f_L(\omega)-1) 
\left[ \left( {\bm G}^{\rm r}_{\Lambda} \frac{d{\bm \Sigma}^{\rm r}_{U}}{d\Lambda}{\bm G}^{\rm r}_{\Lambda} \right.  \right. \right.\nonumber \\
& \left. \left. - \left. {\bm G}^{\rm a}_{\Lambda} \frac{d{\bm \Sigma}^{\rm a}_{U}}{d\Lambda}{\bm G}^{\rm a}_{\Lambda} \right)_{11}(\omega) \right] \left({\bm G}^{\rm r}_{\Lambda} \frac{d{\bm \Sigma}^{\rm r}_{U}}{d\Lambda}{\bm G}^{\rm K}_{\Lambda} \right)_{11}(\omega) \right. \nonumber \\
& \left.
- \left({\bm G}^{\rm K}_{\Lambda} \frac{d{\bm \Sigma}^{\rm a}_{U}}{d\Lambda}{\bm G}^{\rm a}_{\Lambda} \right)_{11}(\omega) \right),
\label{eq:flow eq. of the current}
\end{align}
with initial condition $ \bar{I}_{\Lambda_{\rm init}}=0$. 
As mentioned above, it contains the flowing current-vertex function 
$\bar{{\bm \gamma}}^{(2,1)}_{\Lambda}$. The right-hand side is zero if the infinite hierarchy 
of flow equations is kept, but may become finite if we use approximations, e.g., the 
static one.

Equation (\ref{eq:flow eq. of the current}) together with the expression for the self-energy 
Eq.~(\ref{eq:flow eq. of the self-energy}) as well as the vertex functions 
Eqs.~(\ref{eq:flow eq. of ret. three-point vertex func.}) and 
(\ref{eq:flow eq. of Keldysh three-point vertex func.}) can be solved numerically.
The resulting value of the relative difference, $\left| \bar{I} \right| / I^{\rm MW}$ 
as a function of the dimensionless interaction $u\equiv U/\Delta$ is plotted in 
Fig.~\ref{fig:Diff_Current_vs_U}.
Within the static approximation the difference should be of second order in $u$, i.e., 
$ \bar{I} \equiv \bar{I}_{\Lambda=0} = {\cal O}(u^2)$, which is consistent with the results  
shown in Fig.~\ref{fig:Diff_Current_vs_U} for 
various $V$. This finding indicates that we can consistently 
determine the current by solving its flow equation and that the flow of the current-vertex function was 
properly implemented. As briefly discussed in the next section, which is mainly on the noise, we 
can reproduce all the known results for the current, e.g., power-law scaling with a $U$-dependent 
exponent at large voltages from $I^{\rm flow}_{\Lambda}$.

\subsection{On-resonance current noise}
\label{subsec:noise in the scaling limit}

\begin{figure}[t]
\centering
\includegraphics*[width=0.85\hsize]{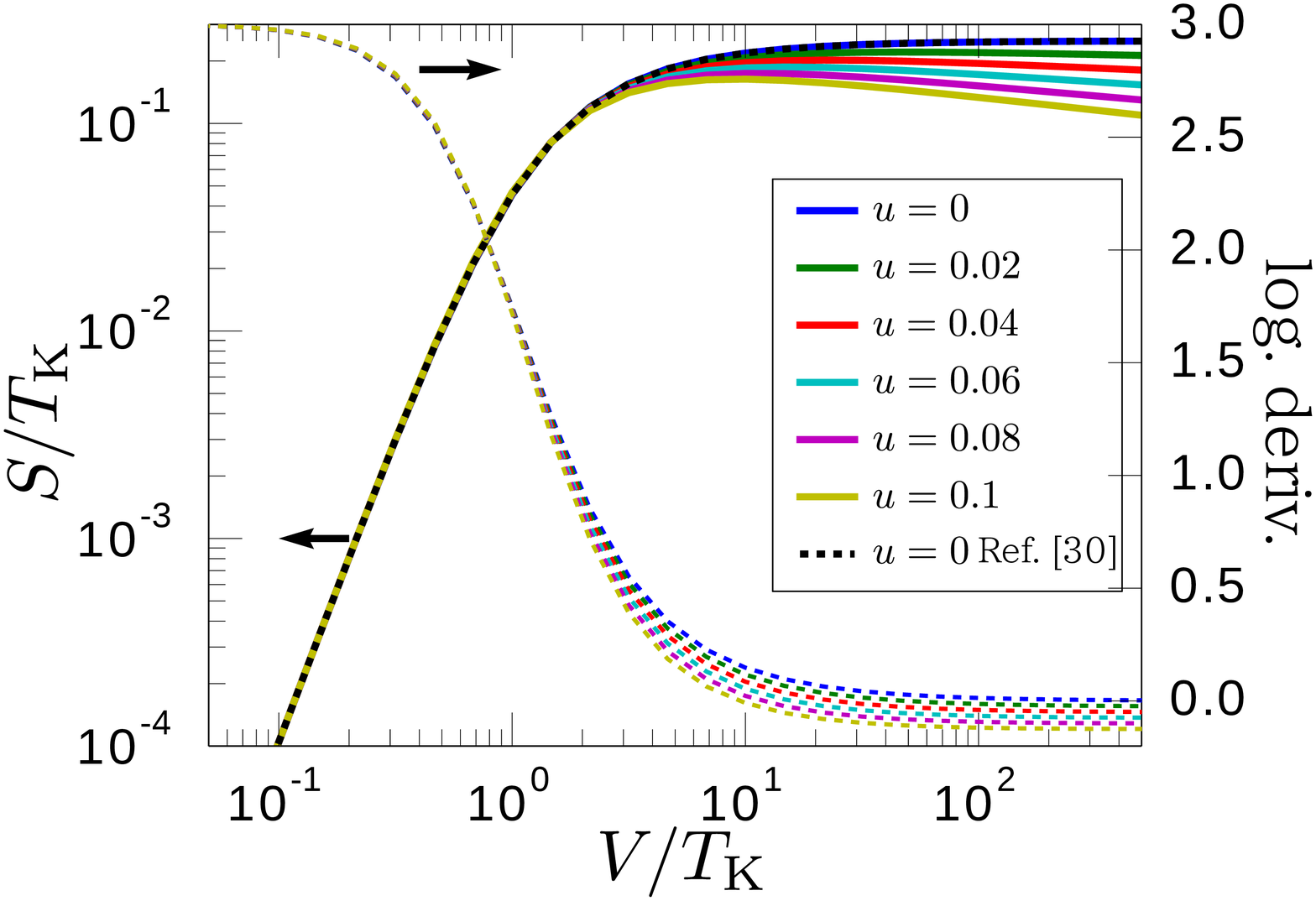}
\caption{
  \label{fig:noise with diff. U}
  (Color online)
  The dependence of noise and its logarithmic derivative on $V$ for various $u$.
  The parameters are $\epsilon / T_{\rm K}=0$, $t / \Delta=0.001$, and $T / T_{\rm K}=0$.
}
\end{figure}

\begin{figure*}[t]
\centering
\includegraphics*[width=0.8\hsize]{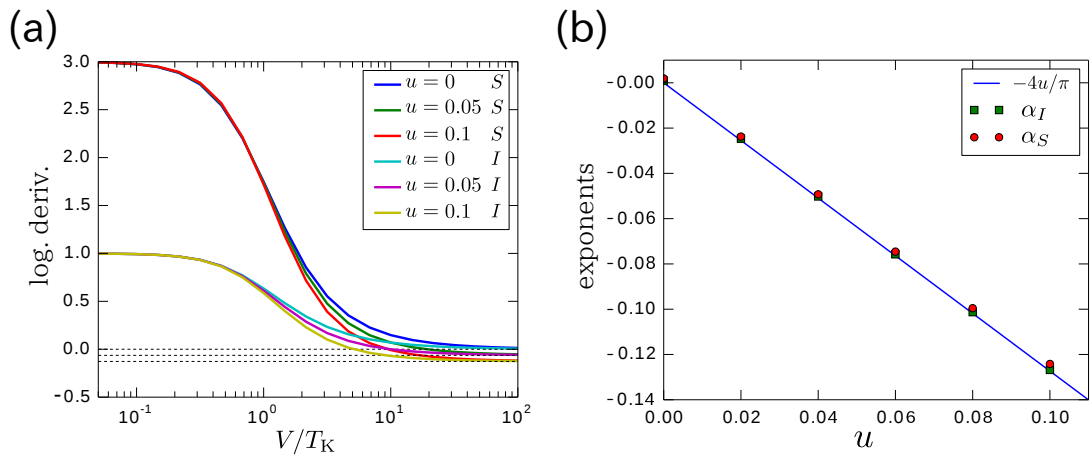}
\caption{
  \label{exponents various U}
  (Color online)
  (a) The dependence of the logarithmic derivative of the current and noise on $V$ for various $u$.
  (b) The exponents at large bias voltages for various $u$.
  The parameters are $\epsilon / T_{\rm K}=0$, $t / \Delta=0.001$, and $T / T_{\rm K}=0$.
}
\end{figure*}

It was established by previous works that the low-energy physics of the IRLM is governed by a single 
energy scale, $T_{\rm K}$ (see, e.g., Ref.~\onlinecite{karrasch2010functional}). Here this 
universal energy scale is introduced as $T_{\rm K} \equiv 8 |\bar{t}^{\rm ren}|^2 / \Delta$ with 
the renormalized hopping amplitude $\bar{t}^{\rm ren} \equiv t +  \left. \bm{\Sigma}^{\rm r}_{12} \right|_{T=V=\epsilon=0}$
at the end of the RG flow. An alternative definition using the susceptibility is discussed in 
Appendix~\ref{app:T_K}.  The current shows a crossover from the linear response regime to  
power-law decay\cite{boulat2008twofold,karrasch2010functional} at $V \simeq T_{\rm K}$.
Hence, it is natural to expect $T_{\rm K}$ as the characteristic energy scale of the current noise 
as well. It is sufficient to focus on the component $ S \equiv S_{LL}(\omega=0)$ due to the charge 
conservation. In this subsection we consider the transport on resonance with 
$\epsilon=0$.

The dependence of the temperature $T=0$ zero-frequency current noise obtained by numerically 
solving the flow equations (for details of the implementation see Appendix~\ref{app:numerical details}) 
on the bias voltage $V$ is shown in Fig.~\ref{fig:noise with diff. U} for various $u$. The 
logarithmic derivative $d \log[S(V)]/d \log(V)$, approximated by centered differences, is shown 
in the same figure. If $S(V)$ is governed by power-law behavior, the exponent can be read off from 
the plateau value of this quantity. From this, it is evident that $S(V)$ is proportional 
to $V^3$ for small $V$. The curves for the current noise collapse into a single one in the 
linear response regime ($V\leq T_{\rm K}$) if properly scaled by $T_{\rm K}$. This indicates 
that the prefactor of the leading $(V/T_{\rm K})^3$ behavior of the noise is independent of 
the two-particle interaction. For $u=0$ the noise computed by FRG agrees with the analytic 
expression given in Ref.~\onlinecite{PhysRevB.82.205414}. The latter is shown as the thick 
dashed line in Fig.~\ref{fig:noise with diff. U}.

It is well established that the $T=0$ current shows power-law suppression at high bias voltages, 
\begin{align}
I / T_{\rm K} & \sim \left( V / T_{\rm K} \right)^{\alpha_I},
\end{align}
with an interaction-dependent exponent which to leading order in $u$ is given by
\begin{align}
\label{eq:exponent of the current}
  \alpha_{I} = -\frac{4u}{\pi}.
\end{align}
The constant logarithmic derivative at large bias voltages in Fig.~\ref{fig:noise with diff. U} 
indicates that the current noise exhibits power-law behavior in the same regime as well:
\begin{align}
S / T_{\rm K} & \sim \left( V / T_{\rm K} \right)^{\alpha_S} .
\end{align}
The logarithmic derivatives of the current and noise are compared in Fig.~\ref{exponents various U}.
The dashed lines in Fig.~\ref{exponents various U}(a) and the solid line in Fig.~\ref{exponents various U}(b) 
indicate the exponent Eq.~(\ref{eq:exponent of the current}). For the current, this value can be 
obtained analytically using FRG.\cite{karrasch2010functional} 
The logarithmic derivative of the noise is found to reach the same value as that of the current at 
sufficiently large bias voltages. 
Previous works employing FRG showed that the behavior of the current can be understood from an 
effective noninteracting model with renormalized parameters, in particular a renormalized
level-lead hopping.\cite{karrasch2010functional} 
As the current-vertex corrections enter the expression for the noise, it is not obvious that
a similar mapping can be used for the noise. 
The contribution of the vertex correction to the current noise is discussed in more detail in the 
next section.

\begin{figure*}[t]
\centering
\includegraphics*[width=0.8\hsize]{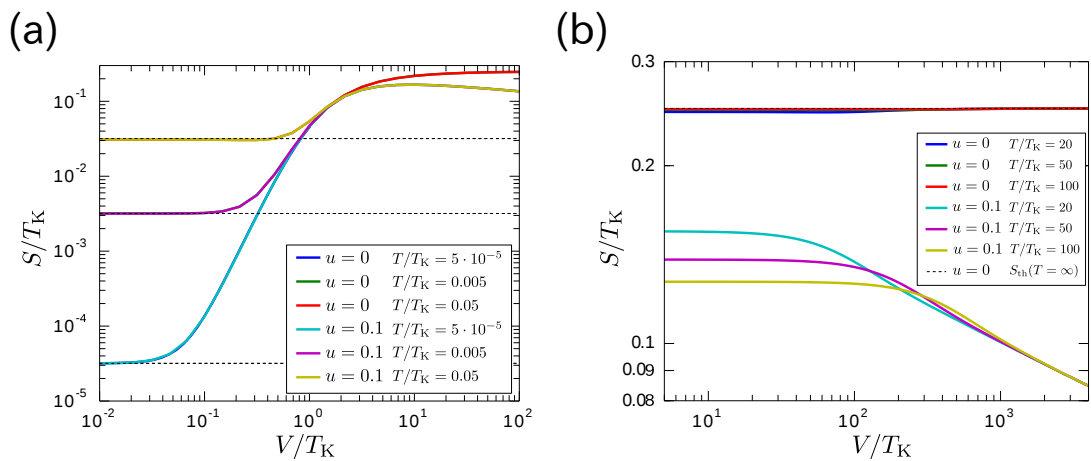}
\caption{
  \label{fig:effect of temperature on noise}
  (Color online)
  The dependence of the noise on $V$ for (a) $T<T_{\rm K}$ and (b) $T>T_{\rm K}$ for various $u$ and $T$.
  The parameters are $\epsilon / T_{\rm K}=0$ and $t / \Delta=0.001$.
}
\end{figure*}

We show plots of the noise as a function of $V$ for various temperatures in 
Fig.~\ref{fig:effect of temperature on noise}. The black dashed lines in Fig.~\ref{fig:effect of temperature on noise}(a) 
are the thermal noise calculated via the fluctuation-dissipation theorem $S_{\rm th}=4 GT$ for each 
temperature with the linear conductance defined (and numerically computed) as 
$G \equiv d \left. I / dV \right|_{V=0}$. 
The excellent agreement confirms that the current noise obeys the fluctuation dissipation relation 
in the zero-bias limit, $S(V/ T_{\rm K} \rightarrow 0)=4GT$. The crossover from thermal  to shot noise 
occurs around voltages which fulfill $S_3V^3 \sim  TG$, where $S_3$ is the coefficient of the $V^3$ 
term in the current noise $S$.
The independence of the coefficient $S_3$ on $u$ in the low-bias regime is discussed later in detail.
The power-law behavior at large voltages is observed for temperatures 
sufficiently lower than $T_{\rm K}$. The current noise calculated at temperatures larger than $T_{\rm K}$ 
is shown in Fig.~\ref{fig:effect of temperature on noise}(b). The power-law decay at high bias voltages survives 
even in this limit if the bias voltage is larger than $T$. In other words, the current noise exhibits 
power-law decay at sufficiently large voltages which satisfy $V \gg \max\{T,T_{\rm K}\}$. Due to this 
renormalization effect, the value of the current noise at high voltages can become even smaller than the 
value in the zero bias limit. We note, however, that the current is suppressed as well with the same exponent.

The dependence of the equilibrium thermal noise on $T$ is shown in Fig.~\ref{fig:FDT}.
Except for the vertex correction, which is irrelevant at small voltages (see below) the thermal 
noise is written as
\begin{align}
  S_{\rm th}= \frac{1}{\pi} \int  d\omega T_{LR}(\omega)
&\left[ f_{L}(\omega)(1-f_L(\omega)) \right. \nonumber \\
    &\left. +f_R(\omega)(1-f_R(\omega))\right],
\end{align}
with
\begin{align}
\label{eq:Transmission amplitude}
  T_{LR}(\omega) \equiv \Delta_L \Delta_R 
\left({\bm G}^{\rm r}\right)_{13}(\omega)\left({\bm G}^{\rm a}\right)_{31}(\omega).
\end{align}
Hence, the numerically observed power-law decay at high temperature in Fig.~\ref{fig:FDT} can be 
understood as a renormalization of the transmission amplitude.
As is shown in the figure, the thermal noise can be exactly translated into the linear conductance 
via the fluctuation-dissipation relation, $4 G T$. This is owing to the reservoir cutoff scheme, in 
which the KMS condition is guaranteed.

\begin{figure}[t]
\centering
\includegraphics*[width=0.85\hsize]{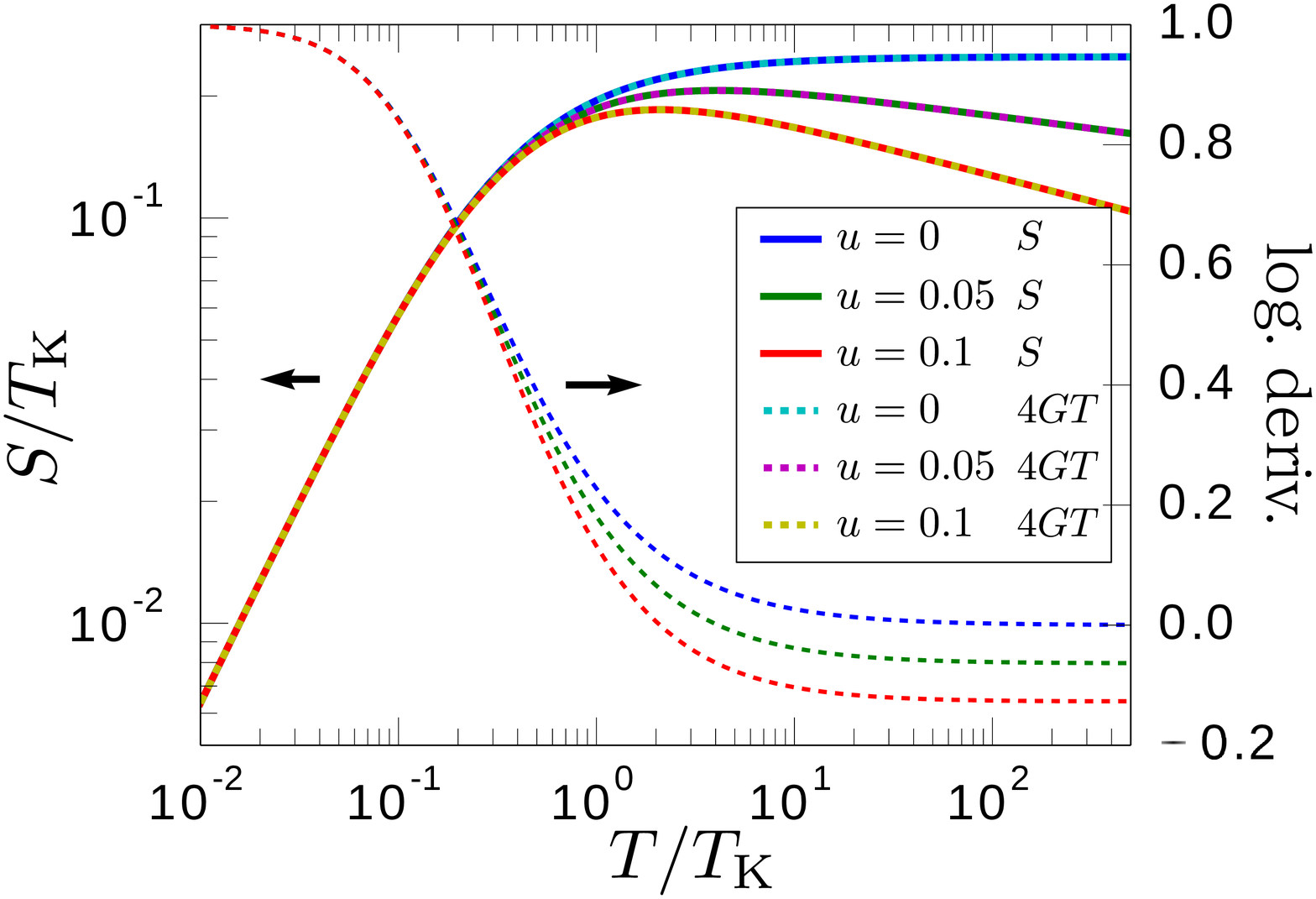}
\caption{
  \label{fig:FDT}
  (Color online)
  The dependence of the equilibrium noise $S$ on $T$ for various $u$.
  The parameters are $\epsilon / T_{\rm K}=0$, $V/T_{\rm K}=0$, and $t / \Delta=0.001$.
}
\end{figure}

\subsection{The effective charge}
\label{subsec:results for effective charge}

\begin{figure}[t]
\centering
\includegraphics*[width=0.75\hsize]{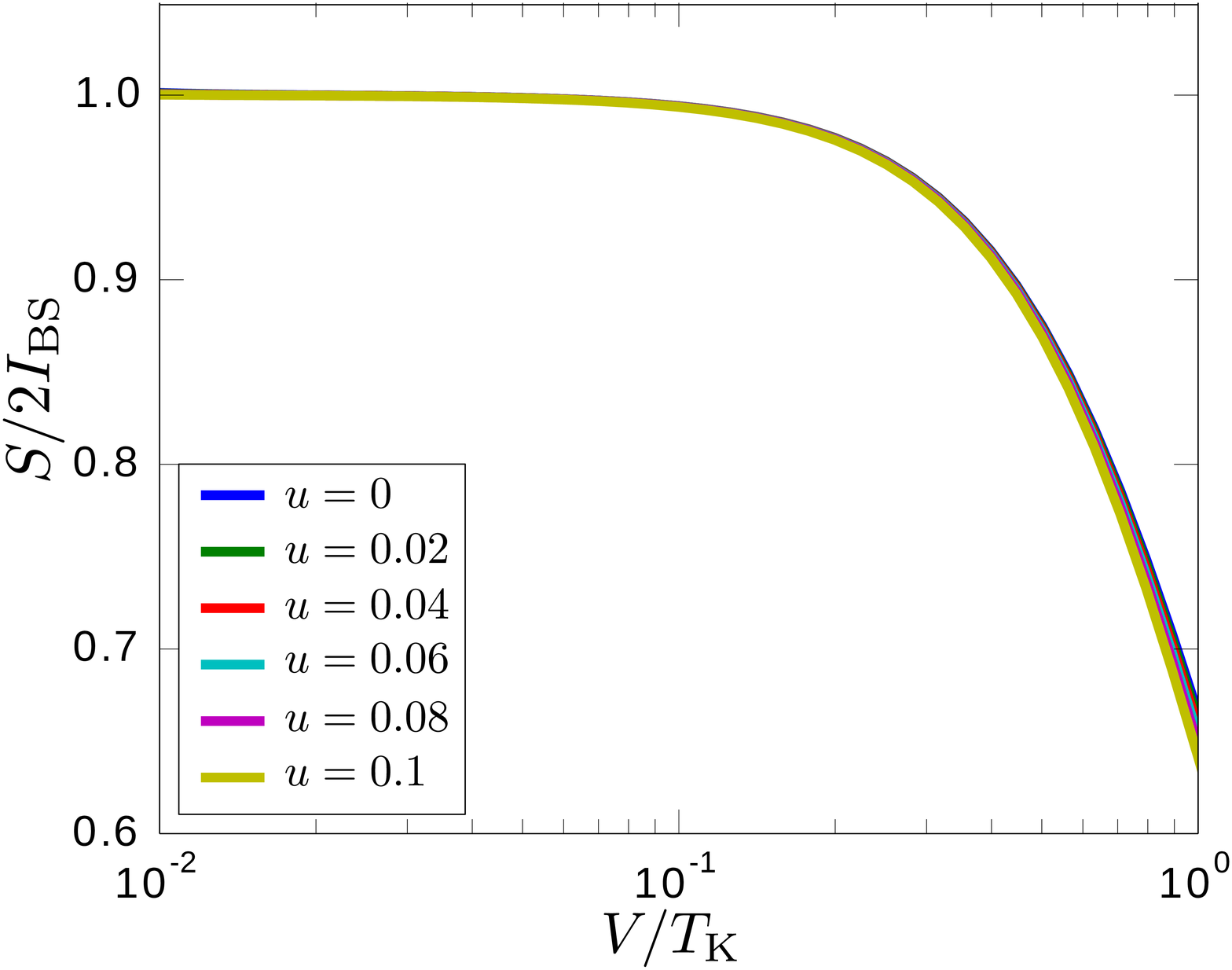}
\caption{
  \label{fig:backward effective charge}
  (Color online)
  The dependence of the ratio between the noise and the backscattering current on $V$ for various $u$.
  The parameters are $\epsilon / T_{\rm K}=0$, $t / \Delta=0.001$, and $T / T_{\rm K}=0$.
}
\end{figure}

The ratio between the noise and the current can be interpreted as an effective charge of carriers 
when the transport is governed by Poisson statistics.\cite{Blanter2001}
On resonance, the effective charge is defined as the ratio between the noise and the 
backscattering current.\cite{branschadel2010shot,carr2011full}
For $u=0$, the IRLM becomes the resonant level model and can be 
solved exactly. In this case, the effective charge $e^*$ is $e$.
Another solvable point is the self-dual one\cite{branschadel2010shot,carr2011full} reached at relatively 
large interaction. At this, field  theoretical techniques and the density-matrix renormalization group 
approach were utilized to show that the effective charge is $2e$. It is, however, unknown, how $e^*$ 
crosses over from $e$ to $2e$ when $u$ is increased. In this subsection, we study this issue using 
FRG. Since our scheme is based on an expansion in terms of the interaction strength (on the right-hand side
of RG flow equations), we are bound to small-to-intermediate $u$.

The effective charge is defined as
\begin{align}
\label{eq:effective charge}
  e^{*} =  \lim_{V\rightarrow 0} \frac{S(V)}{2I_{\rm BS}(V)},
\end{align}
with the backscattering current
\begin{align}
  I_{\rm BS} \equiv GV - I.
\end{align}
The dependence of the ratio $S/2I_{\rm BS}$ on $V$ is shown in 
Fig.~\ref{fig:backward effective charge} for various $u$.
It is evident that the value of $e^*/e$ can be reliably read off at 
$V / T_{\rm K}=10^{-2}$. Obviously $e^*$ does not depend on the interaction in our 
approximation, from which we conclude that $e^* /e =1+{\cal O}\left( u^2 \right)$ 
as all terms to linear order in $u$ are kept in the truncated RG equations.

This numerical observation can be substantiated by analytic considerations.
We first discuss the relation between the effective charge and the 
vertex correction. The transmission amplitude Eq.~(\ref{eq:Transmission amplitude}) can 
be expanded in terms of the bias voltage as
\begin{align}
  T_{\rm LR}(\omega)= T^{(0)}_{\rm LR}(\omega) - T^{(2)}_{\rm LR}(\omega) \left(\frac{V}{T_{\rm K}}\right)^2 + \cdots,
\end{align}
in the linear-response regime ($V<T_{\rm K}$).
If the bias voltage is much smaller than the scale of the energy dependence of the 
transmission amplitude $(V \ll T_{\rm K})$, the current can be evaluated at the Fermi energy as
\begin{align}
  \frac{I}{T_{\rm K}}= \frac{1}{2\pi} \left[ T^{(0)}_{\rm LR} \frac{V}{T_{\rm K}} - T^{(2)}_{\rm LR} \left(\frac{V}{T_{\rm K}}\right)^3  \right].
\end{align}
Since we are considering the on-resonance case, the zeroth-order coefficient $T^{(0)}_{LR}$ 
is unity. The leading term of the backscattering current is thus
\begin{align}
  \frac{I_{\rm BS}}{T_{\rm K}} =  \frac{1}{2\pi}T^{(2)}_{\rm LR}(0)\left(\frac{V}{T_{\rm K}}\right)^3.
\end{align}
We expand the (odd) backscattering current as
\begin{align}
\label{eq: def. of Coef. of I_BS}
  I_{\rm BS}= G_3  \left(\frac{V}{T_{\rm K}}\right)^3 + G_5 \left(\frac{V}{T_{\rm K}}\right)^5 + {\cal O} \left( \left(\frac{V}{T_{\rm K}}\right)^7 \right).
\end{align}
From field theoretical considerations\cite{PhysRevLett.112.216802} it is known that the 
backscattering current is given by
\begin{align}
\label{eq:Coef. of I_BS}
& \frac{2\pi I_{\rm BS} }{ T_{\rm K} } 
=\frac{1}{3} \left[ 1 + {\mathcal O}(u^2) \right] 
\left( \frac{V}{T_{\rm K}}\right)^3  \nonumber \\ & - \frac{1}{5} \left[  1- \frac{20}{3} \frac{u}{\pi} 
+ {\mathcal O}(u^2) \right] \left( \frac{V}{T_{\rm K}}\right)^5 
+ {\mathcal O} \left( \left[ \frac{V}{T_{\rm K}}\right]^7 \right).
\end{align}
This analytical result can also be obtained by FRG  (see endnote [52] of 
Ref.~\onlinecite{PhysRevLett.112.216802};  for a similar analysis of the current as function 
of temperature see Ref.~\onlinecite{PhysRevB.87.075130}). By comparing the coefficients in 
Eq.~(\ref{eq:Coef. of I_BS}) with those in Eq.~(\ref{eq: def. of Coef. of I_BS}), we find
\begin{align}
  \frac{G_3}{T_{\rm K}}=\frac{1}{2\pi}T^{(2)}_{\rm LR}(0) = \frac{1}{6\pi},
\end{align}
independent of $U$. We note in passing that this result can be reproduced by our 
numerical calculations as shown in Fig.~\ref{fig:num. and denom. of eff. charge}(a) 
if $T_{\rm K}$ is properly chosen (see Appendix~\ref{app:T_K}). This exemplifies that our 
numerics gives highly accurate results and that coefficients of expansions in $V/T_{\rm K}$ 
can be reliably determined.  

Considering the above discussion and the definition of $e^*$, the remaining question is whether the 
leading $(V / T_{\rm K})^3$ term in the current noise, $S_3$, has an order $u$ correction or not.
We already mentioned above that this does not seem to be the case [see Fig.~\ref{fig:noise with diff. U}].
To analyze this further, we show the dependence of $S_3$ on bias voltage $V$ in 
Fig.~\ref{fig:num. and denom. of eff. charge}(b). The third-order coefficient $S_3$ is independent 
of the interaction strength 
at low bias voltages, which is consistent with the result of $e^*/e$ shown in 
Fig.~\ref{fig:backward effective charge}.

\begin{figure*}[t]
\centering
\includegraphics*[width=0.8\hsize]{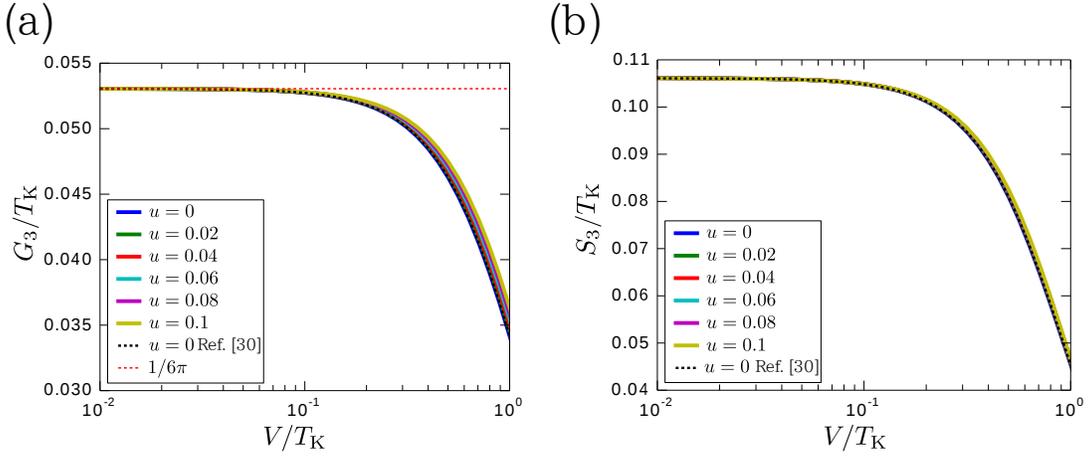}
\caption{
  \label{fig:num. and denom. of eff. charge}
  (Color online)
  (a) The dependence of the third order coefficient of the (a) current $G_3$ and (b) noise $S_3$ on $V$ for various $u$.
  The parameters are $\epsilon / T_{\rm K}=0$, $t / \Delta=0.001$, and $T / T_{\rm K}=0$.
}
\end{figure*}

In the subsequent discussion, we give a microscopic explanation of the $u$ independence of 
the third-order coefficient, $S_3$, by dividing the noise into the bubble term and the vertex 
correction. If we denote the bubble and vertex correction terms by $S_0 \equiv S^{0}_{LL}(\omega=0)$ 
and $S_U \equiv S^{U}_{LL}(\omega=0)$, respectively, the current noise can be written as
\begin{align}
S=S_{0}+S_{U}.
\end{align}
At zero-temperature, the bubble term is given as
\begin{align}
  S_{0}=\frac{1}{\pi} \int^{V/2}_{-V/2} d\omega\; T_{LR}(\omega)\left[1-T_{LR}(\omega)\right].
\end{align}
If we ignore the frequency dependence of the transmission amplitude, we obtain
\begin{align}
  \frac{S_{0}}{T_{\rm K}}= \frac{1}{\pi} \frac{V}{T_{\rm K}} T_{LR}(0)\left[1-T_{LR}(0)\right].
\end{align}
Since we are considering the on-resonance case [$T^{(0)}_{\rm LR}(0)=1$], the  
contribution linear in $V$ to the bubble term vanishes. The lowest order contribution in $V$ is thus
\begin{align}
  \frac{S_{0} }{T_{\rm K}}
  &= \left. \frac{1}{\pi}\left( \frac{V}{T_{\rm K}}\right)^3 T^{(2)}_{\rm LR}(0) \left[ 2T^{(0)}_{\rm LR}(0)-1\right] \right|_{T^{(0)}_{LR}(0)=1} \nonumber \\
  &= \frac{1}{\pi}\left( \frac{V}{T_{\rm K}}\right)^3 T^{(2)}_{\rm LR}(0) .
\end{align}
With this, the effective charge is obtained as
\begin{align}
  e^{*}
  &=  \left. \frac{S}{2I_{\rm BS}} \right|_{V=0} \nonumber \\
  &= 1 + \left. \frac{S_U}{2I_{\rm BS}} \right|_{V=0} \nonumber \\
  &= 1 + \left.  \frac{3\pi S_U}{\left( V / T_{\rm K}\right)^{3}}  \right|_{V=0}.
\label{eq:effective charge and the vertex correction}
\end{align}
This relation shows that the $U$ dependence of the effective charge is incorporated 
via the vertex correction $S_U$ analyzed next.

\begin{figure}[t]
\centering
\includegraphics*[width=0.8\hsize]{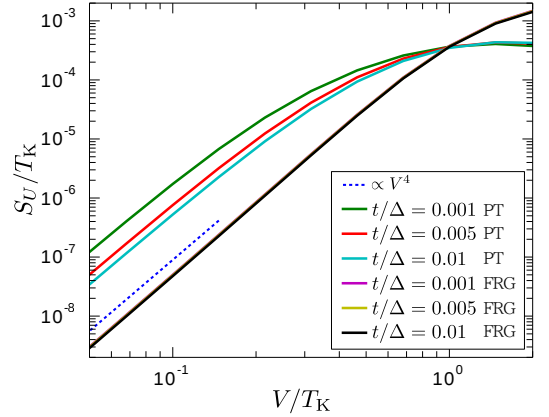}
\caption{
  \label{fig:vertex correction FRG VS PT}
  (Color online)
  The vertex correction calculated using a FRG scheme and a plain perturbation theory for various $t$ as a function of $V$.
  The parameters are $u=0.1$, $\epsilon / T_{\rm K}=0$, and $T / T_{\rm K}=0$.
}
\end{figure}

The current-vertex functions enter the expression for the vertex correction 
$S_U$ [see Eq.~(\ref{eq:expression of vertex correction})].  The three-point vertex function 
can be computed in two different ways by either plain perturbation theory or by solving its 
flow equations (\ref{eq:flow eq. of ret. three-point vertex func.}) and 
(\ref{eq:flow eq. of Keldysh three-point vertex func.}). It is well established that 
the self-energy computed in leading order perturbation theory in $u$ is plagued by a logarithmically divergent
term.\cite{karrasch2010functional} 
To avoid this known problem in a perturbative computation of $S_U$, depicted 
in the right diagram of Fig.~\ref{fig:Diagram of Current noise}, we dressed the two 
propagators by the self-energy computed within FRG. This way we single out possible problems 
of a perturbative calculation of the three-point vertex function itself.  
The dependence of $S_U$ on $V$ obtained by perturbation theory and FRG is 
shown in Fig.~\ref{fig:vertex correction FRG VS PT} 
for different $t/\Delta$. In both computations, the vertex correction scales as $V^4$ 
for small $V$, which is subleading compared with the bubble term which goes as $V^3$.
A significant difference is that the results obtained by perturbation theory becomes gradually 
larger the deeper one goes into the scaling limit $t /\Delta \ll 1$. 
The vertex corrections calculated using FRG for the three-point vertex are free of 
this problem and collapse into a single curve if rescaled by $T_{\rm K}$. This indicates 
that, in analogy to the self-energy, the FRG regularizes the divergences of the vertex correction.

\begin{figure}[t]
\centering
\includegraphics*[width=0.8\hsize]{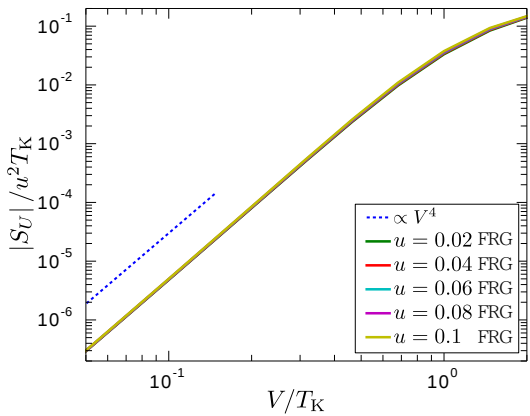}
\caption{
  \label{fig:VC scaled with u^2 for various U whole V}
  (Color online)
  The vertex correction divided by $u^2$ calculated using our FRG scheme for various $u$ as a function of $V$.
  The parameters are $\epsilon / T_{\rm K}=0$, $t/\Delta=0.001$, and $T / T_{\rm K}=0$.
}
\end{figure}

The vertex correction fully computed by FRG and divided by $u^2$ is plotted for various $u$ in 
Fig.~\ref{fig:VC scaled with u^2 for various U whole V}.
From this figure, it is evident that $S_U$ depends on $u$ and 
$V$ as $\left| S_U \right|/ T_{\rm K} \propto u^2 (V/T_{\rm K})^4$.
Because of the $u^2$ prefactor in the linear response regime ($V<T_{\rm K}$), the vertex 
correction is not under control within the static approximation which only contains all terms 
to order $u$. However, the vertex correction does not contribute to the effective charge because 
it is of order $V^4$ while the bubble term scales as $V^3$ 
[see Eq.~(\ref{eq:effective charge and the vertex correction})].

\subsection{Off-resonance current noise}
\label{subsec:particle-hole asymmetry}

\begin{figure*}[t]
\centering
\includegraphics*[width=0.8\hsize]{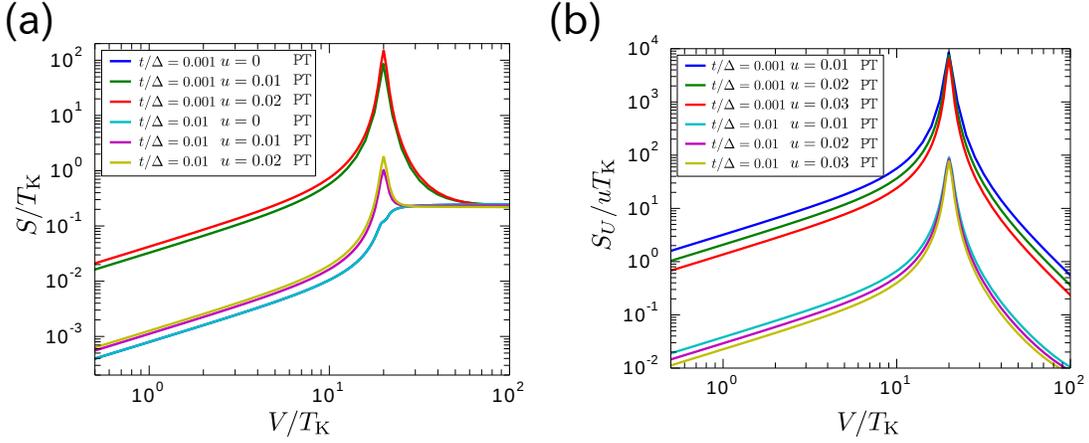}
\caption{
  \label{fig:curret noise for p-h asymmetric case PT}
    (Color online)
  (a) The current noise and (b) the vertex correction calculated from a plain perturbation theory for various $u$ and $t$ away from the particle-hole symmetric point as a function of $V$.
  The parameters are $\epsilon / T_{\rm K}=10$ and $T / T_{\rm K}=0$.
}
\end{figure*}

In this subsection, we investigate the current noise away from the particle-hole symmetric point 
($\epsilon \neq 0$).
We start out by considering the current noise with the three-point vertex functions calculated using plain 
perturbation theory, but with all propagators dressed by the FRG self-energy [see above].
The noise as a function of voltage for different $u$ and $t / \Delta$ 
is shown in Fig.~\ref{fig:curret noise for p-h asymmetric case PT}(a).
If the level energy aligns with one of the leads chemical potentials 
($\epsilon \sim \pm V/2$) a peak develops.
The peak exhibits divergent behavior for decreasing $t/\Delta$, that is, when going into the scaling 
limit, at fixed $u$. We reemphasize that this divergence originates from the vertex function, as 
logarithmic divergences of the self-energy have already been removed by employing the FRG self-energy. 
The vertex correction to the noise $S_U$ divided by $u$ is shown in
Fig.~\ref{fig:curret noise for p-h asymmetric case PT}(b). 
From this we conclude that the term diverging for $t/\Delta \to 0$ has a prefactor $u$. Plain 
perturbation theory can thus not be used to study the current noise away from particle-hole symmetry 
in the scaling limit even for very small $u$.

\begin{figure*}[t]
\centering
\includegraphics*[width=0.8\hsize]{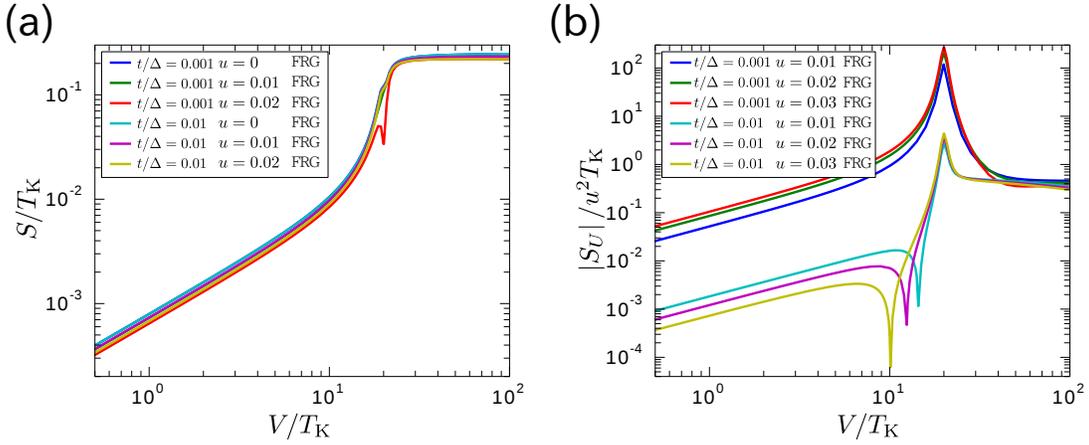}
\caption{
  \label{fig:curret noise for p-h asymmetric case with diff. U diff. t}
  (Color online)
  (a) The current noise and (b) the vertex correction divided by $u^2$ calculated using our FRG scheme for various $u$ and $t$ away from the particle-hole symmetric point as a function of $V$.
    The parameters are $\epsilon / T_{\rm K}=10$ and $T / T_{\rm K}=0$.
}
\end{figure*}

The current noise and its vertex correction determined by our FRG scheme are shown in 
Fig.~\ref{fig:curret noise for p-h asymmetric case with diff. U diff. t}. The vertex functions are 
obtained by solving their flow equations Eqs.~(\ref{eq:flow eq. of ret. three-point vertex func.}) 
and (\ref{eq:flow eq. of Keldysh three-point vertex func.}). The divergent behavior of the current noise 
observed in Fig.~\ref{fig:curret noise for p-h asymmetric case PT}(a) is essentially 
removed for the curves in Fig.~\ref{fig:curret noise for p-h asymmetric case with diff. U diff. t}(a).
Further down we comment on the weak features still visible in the regime $\epsilon \sim \pm V/2$.
This indicates that, as for the self-energy (and thus the current when employing the Meir-Wingreen formula), 
the RG-based scheme regularizes the leading-order divergences. To further analyze this the vertex 
correction to the noise $S_U$ divided by $u^2$ is shown in 
Fig.~\ref{fig:curret noise for p-h asymmetric case with diff. U diff. t}(b). This figure indicates that 
the divergence with prefactor $u$ of first order perturbation theory 
(see Fig.~\ref{fig:curret noise for p-h asymmetric case PT}(b)) is pushed to order $u^2$ within FRG. 
As our truncation does not contain all terms ${\mathcal O}(u^2)$ we do not 
control $S_U$ to this order. This second order divergence in $S_U$ manifests as the artificial dip of 
the noise for $u=0.02$ and $t/\Delta=0.001$ and the shoulders for the other parameter sets 
found in Fig.~\ref{fig:curret noise for p-h asymmetric case with diff. U diff. t}(a).
When next considering larger interactions we thus take $t/\Delta=0.01$ instead of $0.001$ as 
before to avoid this order $u^2$ artifact. 

\begin{figure}[t]
\centering
\includegraphics*[width=0.75\hsize]{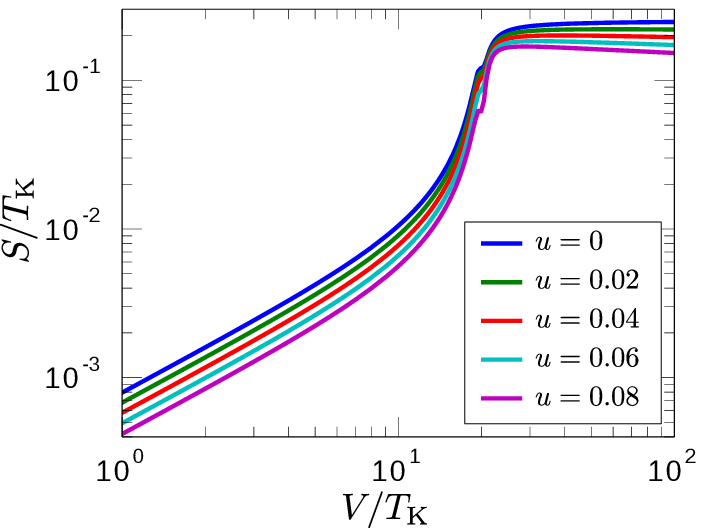}
\caption{
  \label{fig:Noise away from p-h sym.}
  (Color online)
  The dependence of the current noise on $V$ for various $u$ away from the particle-hole symmetric point.
  The parameters are $\epsilon / T_{\rm K}=10$, $t / \Delta=0.01$, and $T / T_{\rm K}=0$.
}
\end{figure}

The current noise as a function of $V$ is shown in 
Fig.~\ref{fig:Noise away from p-h sym.} for a variety of $u$. 
The current noise is proportional to $V$ in the linear-response regime
(see the plot for $V<T_{\rm K}$ as well as the analytic considerations in 
Sec.~\ref{subsec:results for effective charge}). 
At large bias voltages, the current crosses overs to a power-law decay with an interaction-dependent exponent.
The exponent agrees with the one found for $\epsilon=0$: $\alpha_S=-4u/\pi$.  
This is in accordance with our intuition that 
the bias voltage dominates the transport for $V \gg \epsilon$. 

\begin{figure}[t]
\centering
\includegraphics*[width=0.85\hsize]{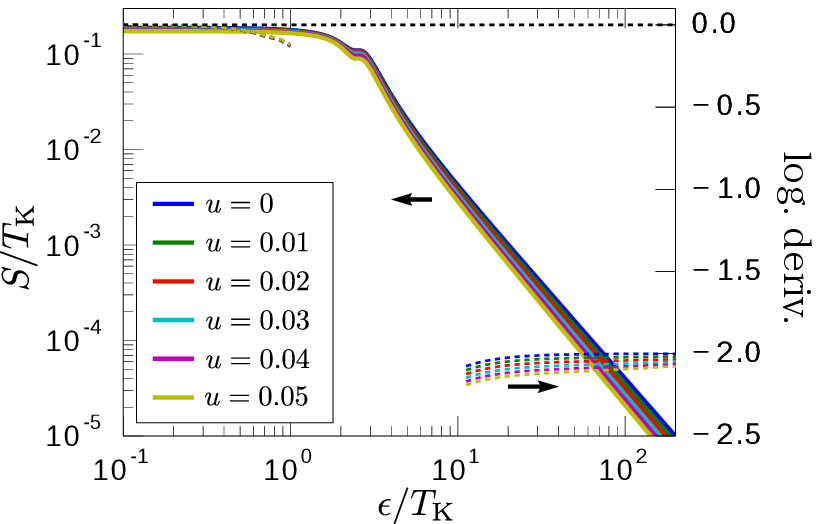}
\caption{
  \label{fig:Noise away from p-h sym. epsilon dep.}
  (Color online)
  The dependence of the current noise and its logarithmic derivative on $\epsilon$ for various $u$.
  The parameters are $V / T_{\rm K}=5$, $t / \Delta=0.01$, and $T / T_{\rm K}=0$.
}
\end{figure}

The dependence of the current noise on $\epsilon$ at fixed $V$ is shown in 
Fig.~\ref{fig:Noise away from p-h sym. epsilon dep.}.
It is independent of $\epsilon$ for $\epsilon \ll V$ as the level is placed inside the bias window.
The noise starts to decrease when the 
level energy is beyond the bias window ($\epsilon \gtrsim V/2$). The weak features found when 
the level is aligned with one of the lead chemical potentials were discussed above.  
For large $\epsilon \gg T_{\rm K}$, the noise crosses over to a power-law decay as a function 
of $\epsilon$ as the renormalization of the hoping amplitude is cut by the level position in this case.
For $u=0$ the exponent is $-2$ and the interacting part of the exponent is found to be twice that 
of the $V$ dependence; see the logarithmic derivative shown in 
Fig.~\ref{fig:Noise away from p-h sym. epsilon dep.} for $\epsilon/T_{\rm K} \leq 1$ and 
$\epsilon/T_{\rm K} \geq 10$.

\section{Summary}
\label{sec:summary}

In the present paper, we have developed a FRG scheme to describe the current noise of the nonequilibrium IRLM. 
The coupled set of flow equations of the current-vertex functions and the self-energy are derived and solved 
to determine the current noise within the lowest-order approximation in the two-particle interaction. 

The vertex correction of the current noise shows divergent behavior in the scaling limit, if it is calculated 
using plain perturbation theory. This divergence is removed in our FRG method at the particle-hole 
symmetric point, which makes it possible to perform a reliable analysis in the deep scaling limit.
In this regime, the current noise is found to show a power-law decay at high voltages characterized by 
the same exponent as that of the current. This property is robust against temperature.
The effective charge of the IRLM at the particle-hole symmetric point can be reliably extracted
and is found to be interaction independent to linear order. This behavior can be understood from 
the properties of the vertex contribution to the noise by combining analytical arguments 
and the numerical results.

The current noise away from the particle-hole symmetric point determined by plain perturbation 
theory shows a severe leading order divergence, which originates from the current-vertex correction.
We showed that the divergent term which is proportional to $u$ is consistently removed in our scheme 
and pushed to order $u^2$; this lies beyond our control. Although the remaining order $u^2$ divergence 
is an obstacle to calculate the current noise for the particle-hole asymmetric case in the scaling limit 
($t /\Delta \ll 1$), we obtain reliable results down to $t/\Delta=0.01$.
We showed that the current noise shows a power-law decay for $\max\{V,\epsilon\} \gg T_{\rm K}$.

The present paper shows that the FRG method allows one to reliably calculate the current noise 
in the scaling regime. A higher order FRG calculation -- possible in principle, complicated in practice --  
would be desirable to further elucidate the crossover of the effective charge from the noninteracting 
case ($e^{*}/e =1$) to the self-dual point ($e^{*}/e =2$) at relatively large $u$.
Furthermore, the higher order contributions need to be taken into account in order to remove a diverging 
term of order $u^2$ away from particle-hole symmetry and discuss the current noise in the deep scaling 
limit for this case. 
Another step for the future would be to extend the FRG treatment 
to determine the full counting statistics of interacting fermion systems.

\begin{acknowledgments}
  We thank Takeo Kato, Akinori Nishino, Katharina Eissing, and Peter Schmitteckert for very useful discussions.
  T.J.S. acknowledges financial support provided by the Advanced Leading Graduate Course 
  for Photon Science (ALPS). This work was supported by the Deutsche Forschungsgemeinschaf via RTG 1995.  
\end{acknowledgments}

\appendix

\section{Initial conditions}
\label{app:Initial conditions}

The initial condition of the self-energy for $\Lambda_{\rm init} \to \infty$ 
is written as
\begin{align}
\left(\bm{\Sigma}^{\rm r}_{U,\Lambda_{\rm init}}\right)_{11}(\omega) &= U_1 n_2,\\
\left(\bm{\Sigma}^{\rm r}_{U,\Lambda_{\rm init}}\right)_{22}(\omega) &= U_1 n_1 + U_3 n_3 ,\\
\left(\bm{\Sigma}^{\rm r}_{U,\Lambda_{\rm init}}\right)_{33}(\omega) &= U_3 n_2,\\
\left(\bm{\Sigma}^{\rm K}_{U,\Lambda_{\rm init}}\right)_{ij}(\omega) &= 0,
\end{align}
where $n_i$ is the occupation of the $i$th site.

By a simple diagrammatic argument,\cite{PhysRevB.81.195109} current-vertex functions are found to 
be identical to those of the noninteracting system in the limit of $\Lambda_{\rm init} \rightarrow \infty$;
\begin{align}
&\left(\bm{\gamma}^{(2,n)}_{\Lambda_{\rm init}}\right)^{\nu'_1\nu_1;{\rm s}\cdots {\rm s}}_{ij;\alpha_1\cdots\alpha_n} (t'_1,t_1;t''_1\cdots t''_n) \nonumber \\
&=\left(\bm{\gamma}^{(2,n)}_{\rm res}\right)^{\nu'_1\nu_1;{\rm s}\cdots {\rm s}}_{ij;\alpha_1\cdots\alpha_n} (t'_1,t_1;t''_1\cdots t''_n) \ \ \ ({\rm for} \ n>0) .
\end{align}
The noninteracting current-vertex functions can be determined using the Ward-Takahashi identity
\begin{align}
&\left(\bm{\gamma}^{(2,1)}_{\rm res}\right)_{11;L} (\tau',\tau;\tau'') \nonumber \\
&= i \left[ \delta(\tau',\tau'') - \delta(\tau,\tau'') \right] 
\left(\bm{\Sigma}_{\rm res}\right)_{11} (\tau',\tau).
\end{align}
The other components of the three-point vertex functions are zero because the source 
field $A_L(\tau)$ is only included in the (1,1)-component of the tunneling self-energy 
Eq.~(\ref{eq:tunneling self-energy for IRLM}).
The initial conditions of the three-point current-vertex functions are obtained as
\begin{align}
\left(\bm{\gamma}^{(2,1)}_{\Lambda_{\rm init}}\right)^{{\rm r};{\rm s}}_{ij;\alpha_1} (\omega_1,\omega_1;0) 
&= - \delta_{i1}\delta_{j1}\delta_{\alpha_1 L} \Delta_L (1-2f_L(\omega_1)),\\
\left(\bm{\gamma}^{(2,1)}_{\Lambda_{\rm init}}\right)^{{\rm a};{\rm s}}_{ij;\alpha_1} (\omega_1,\omega_1;0) 
&=  \delta_{i1}\delta_{j1}\delta_{\alpha_1 L} \Delta_L (1-2f_L(\omega_1)),\\
\left(\bm{\gamma}^{(2,1)}_{\Lambda_{\rm init}}\right)^{{\rm K};{\rm s}}_{ij;\alpha_1} (\omega_1,\omega_1;0) 
&=  - \delta_{i1}\delta_{j1}\delta_{\alpha_1 L} \Delta_L ,\\
\left(\bm{\gamma}^{(2,1)}_{\Lambda_{\rm init}}\right)^{{\tilde {\rm K}};{\rm s}}_{ij;\alpha_1} (\omega_1,\omega_1;0) 
&= \delta_{i1}\delta_{j1}\delta_{\alpha_1 L} \Delta_L.
\end{align}
Here, we show only the case with $\omega_1=\omega'_1$ because we focus on the zero-frequency current 
noise in this paper. We note that $\left(\bm{\gamma}^{(2,1)}_{\Lambda_{\rm init}}\right)^{{\tilde {\rm K}};{\rm s}}$ 
does not need to be zero. The multi-point current vertices are determined by recursively using the Ward-Takahashi 
identity, and the initial conditions for four-point current-vertex functions are
\begin{align}
&\left(\bm{\gamma}^{(2,2)}_{\Lambda_{\rm init}}\right)^{{\rm r};{\rm ss}}_{ij;\alpha_1\alpha_2} (\omega_1,\omega_1;0,0)
= 2i \delta_{i1}\delta_{j1}\delta_{\alpha_1 L}\delta_{\alpha_2 L} \Delta_L ,\\
&\left(\bm{\gamma}^{(2,2)}_{\Lambda_{\rm init}}\right)^{{\rm a};{\rm ss}}_{ij;\alpha_1\alpha_2} (\omega_1,\omega_1;0,0) 
=  -2i  \delta_{i1}\delta_{j1}\delta_{\alpha_1 L}\delta_{\alpha_2 L} \Delta_L ,\\
&\left(\bm{\gamma}^{(2,2)}_{\Lambda_{\rm init}}\right)^{{\rm K};{\rm ss}}_{ij;\alpha_1\alpha_2} (\omega_1,\omega_1;0,0) \nonumber \\
&=  2i  \delta_{i1}\delta_{j1}\delta_{\alpha_1 L}\delta_{\alpha_2 L} \Delta_L (1-2f_L(\omega_1)),\\
&\left(\bm{\gamma}^{(2,2)}_{\Lambda_{\rm init}}\right)^{{\tilde {\rm K}};{\rm ss}}_{ij;\alpha_1\alpha_2} (\omega_1,\omega_1;0,0) \nonumber \\
&= - 2i  \delta_{i1}\delta_{j1}\delta_{\alpha_1 L}\delta_{\alpha_2 L} \Delta_L (1-2f_L(\omega_1)).
\end{align}

The initial conditions of the four-point and higher-point vertex functions are determined by the 
bare action. If we denote the antisymmetrized bare two-particle interaction\cite{negele1988quantum} 
by $U_{ij;kl}$, these vertex functions are written as
\begin{align}
  &\left(\bm{\gamma}^{(4,0)}_{\Lambda_{\rm init}}\right)^{\nu'_1\nu'_2;\nu_1\nu_2}_{ij;kl}(\omega'_1,\omega'_2;\omega_1,\omega'_1+\omega'_2-\omega_1) \nonumber \\ 
 &=
  \left\{
  \begin{array}{l}
    \displaystyle -\nu'_1{U_{ij;kl}} \ \;\;\;{\rm if} \ \nu'_1=\nu'_2=\nu_1=\nu_2, \\        
    0 \ \;\;\;\;\;\;\;\;\;\; {\rm otherwise}.
  \end{array}
  \right. \\
  &\left(\bm{\gamma}^{(4,m)}_{\Lambda_{\rm init}}\right)^{\nu'_1\nu'_2;\nu_1\nu_2;\nu''_1 \cdots \nu''_m}_{ij;kl;\alpha_1\cdots\alpha_m}(\omega'_1,\omega'_2;\omega_1,\omega_2;\omega''_1,\cdots,\omega''_m) \nonumber \\
  &=0 \ \ (m>0).
\end{align}

\section{Numerical details}
\label{app:numerical details}

For the plots in the main text, we take the error of solving the set of 
ordinary differential equations to be $10^{-6} T_{\rm K}$.
The frequency grid points are determined using geometric sequences with a scale factor $\Delta \omega=10^{-8}  T_{\rm K}$.
The number of the grid points is $N_{\rm grid}=4801$, which is sufficient 
to produce $N_{\rm grid}$ independent results.
As our main interest lies in the scaling regime, the band width $\Delta$ should be taken to be large enough for the $t/\Delta$ correction to be negligible.
We used $\Delta=10^4$ and $10^6$ for $t/\Delta=0.01$ and $0.001$, respectively.

\section{Definition of $T_{\rm K}$}
\label{app:T_K}

Several ways exist to define the emergent low-energy scale $T_{\rm K}$. Within the FRG approach the most
natural ones are either by the renormalized hopping amplitude or the susceptibility.
We used the renormalized hopping amplitude in the main text. The other definition utilized in previous FRG 
works is 
\begin{align}
  T^{\rm sus}_{\rm K} \equiv -\frac{2}{\pi} \left( \left. \frac{d \langle n_2 \rangle}{d \epsilon} \right|_{T=V=\epsilon=0} \right)^{-1},
\end{align}
where $\langle n_2 \rangle$ is the occupation of the quantum dot site 2.
Deep in the scaling regime both definitions can equivalently be used when comparing with 
field theoretical results obtained for $t/\Delta \to 0$. We found that, for the $t/\Delta$ reachable
by us, results rescaled with the $T_{\rm K}$ derived from the renormalized hopping show weaker 
$t/\Delta$ corrections and are thus closer to the field theoretical predictions. For this reason 
we used this definition in the main text.


\bibliography{reference_modified}

\end{document}